\documentclass[]{aa}
\usepackage[utf8]{inputenc}
\usepackage{graphicx}
\usepackage{natbib}

\usepackage{array}
\usepackage{makecell}

\usepackage{hyperref}
\hypersetup{
    colorlinks=true,
    linkcolor=blue,
    citecolor=blue,
    urlcolor=blue,
}

\usepackage{float}
\usepackage{xcolor}
\usepackage{mathrsfs }

\usepackage[normalem]{ulem} 

\usepackage{balance}
\usepackage[toc,page]{appendix}

\begin{document}

\title{Gravitational instability of solar prominence threads} 
\subtitle{I. Curved magnetic fields without dips}

\titlerunning{I. Gravitational instability of prominence threads }
\authorrunning{Adrover-Gonz\'alez et al.}

\author{Adrover-Gonz\'alez$^{1,3}$, A., Terradas$^{1,3}$, J., Oliver$^{1,3}$, R., Carbonell$^{2,3}$, M.}

\institute{$^1$Departament de F\'\i sica, Universitat de les Illes Balears (UIB),
E-07122, Spain\\ 
$^2$Departament de Ci\`encies Matem\`atiques i Inform\`atica, Universitat de les Illes Balears (UIB),
E-07122, Spain
\\  $^3$Institute of Applied Computing \& Community Code (IAC$^3$),
UIB, Spain
\\
\email{a.adrover@uib.es}
}

\date{}

\abstract{Prominence threads are dense and cold structures lying on curved magnetic fields that can be suspended in the solar atmosphere against gravity.} {The gravitational stability of threads, in the absence of non-ideal effects, is comprehensively investigated in the present work by means of an elementary but effective model.} {Based on purely hydrodynamic equations in one spatial dimension and applying line-tying conditions at the footpoints of the magnetic field lines, we derive analytical expressions for the different feasible equilibria ($s_{\rm e}$) and the corresponding frequencies of oscillation ($\omega$).} {We find that the system allows for stable and unstable equilibrium solutions subject to the initial position of the thread ($s_0$), its density contrast ($\rho_{\rm t}$) and length ($l_{\rm t}$),  and the total length of the magnetic field lines ($L$). The transition between the two types of solutions is produced at specific bifurcation points that have been determined analytically in some particular cases. When the thread is initially at the top of the concave magnetic field, that is at the apex, we find a supercritical pitchfork bifurcation, while for a shifted initial thread position with respect to this point the symmetry is broken and the system is characterised by an S-shaped bifurcation.} {The plain results presented in this  paper shed new light on the behaviour of threads in curved magnetic fields under the presence of gravity and help to interpret more complex numerical magnetohydrodynamics (MHD) simulations about similar structures.}

\keywords{Magnetohydrodynamics (MHD) --- waves --- Sun: magnetic fields}

\maketitle

\section{Introduction}\label{intro}

The solar atmosphere shows a variety of dynamic structures that are much denser than the background corona. These formations, guided by the curved magnetic field and most likely created by thermal instabilities, have their own dynamics and are undoubtedly affected by the gravity force. Examples of such configurations are solar prominences or coronal rain among others. 

A question that immediately arises regarding these structures, and especially for solar prominences, is how they can be suspended against gravity above the photosphere for long periods. This has led to the development of theoretical prominence models that include magnetic dips where the gravity force projected along the field lines is zero. Analytical examples of such equilibrium configurations were developed some years ago by \citet{hoodanzer90}, \citet{fiedlerhood1992}, and \citet{debruynehood93}. Later, equilibrium models in flux rope configurations were obtained by \citet{lowzhang2004}, \citet{petrietal2007}, and \citet{bloklandkeppens2011}. More recently, numerical equilibria are achieved through relaxation processes by solving the time dependent magnetohydrodynamics (MHD) equations in two-dimensional (2D) configurations \citep{hilliervan2013,terradasetal2013,lunaetal2016}. Three-dimensional (3D) stable magnetic configurations with dips are numerically investigated by \citet{terradasetal2015,terradasetal2016} and \citet{adroverterradas2020}. In these last works the presence of dips is not crucial for the stability of the prominence. However, it seems  that there is the common idea in the solar community that threads or blobs automatically fall along the field lines because of gravity force and that if threads are suspended it is because of magnetic dips. However, from the force-balance point of view, the gas pressure might still support a short segment of  prominence as numerically illustrated by \citet{kohutovaver2017b} and \citet{macga2001}.

A characteristic of the \citet{hoodanzer90} model is that the equilibrium is in general unstable to lateral displacements, meaning that the dense and cold part of the prominence body tends to fall along magnetic field lines essentially owing to the gravity force \citep[see also the numerical experiments of][]{kohutovaver2017b}. This was demonstrated in the stability analysis of \citet{debruynehood93}. Therefore, the presence of magnetic dips in the configuration is not a warranty for stability. On the contrary, in the magnetic configuration studied numerically in  \citet{terradasetal2013} the gravitational instability reported in the \citet{hoodanzer90} model was clearly not taking place.
This is the main motivation of part {\rm I} of the present work, namely to understand the reasons for gravitational instability in a global sense, and not necessarily under the presence of magnetic dips. The effect of magnetic dips will be addressed in part {\rm II}.

At first glance, this instability may seem to be directly related to magnetic buoyancy, which is a phenomenon that has been investigated in great detail by \citet{parker1979}. But it turns out that magnetic buoyancy is not related to this kind of process since it also appears in the absence of fluctuations in the magnetic field so the instability is purely gravitational.

In this work we address gravitational instability using a very simple configuration with the consideration of gravity. This allows us to disentangle the main physical processes that occur in the system. We consider a circular magnetic flux tube in which gravity has a spatially varying parallel component along the field lines. We use elementary physics describing a plasma in this configuration, that is the interplay between gas pressure and gravity force. The magnetic field is assumed not to change as a result of the presence of the density enhancement or thread and  simply acts as a guide for longitudinal motions. We do not include non-ideal effects, which are most likely important in the solar corona (and chromosphere), because in our opinion the purely mechanical situation under line-tying conditions in the absence of dissipative mechanisms has received little attention in the literature and needs to be understood more deeply. 

\section{Background equilibrium solution}\label{background}
We start with the description of our model. We assume that the shape of the magnetic field is circular and that its cross-section, $A_0$, is constant along the field (see dotted curves in Fig.~\ref{figequil} top panel). We denote by $s$ the coordinate along the tube, starting from the left foot. The total length of the magnetic tube is $L$ and the top or apex is located at $s=L/2$. Gravity is pointing downwards and the projection along the field lines in this circular configuration is
given by\begin{eqnarray}\label{gpar}
g_\parallel(s)=-g\, \cos\left(\pi \frac{s}{L}\right),
\end{eqnarray}
where the radius of the field line is $R=L/\pi$.

The hydrostatic equation along the tube is simply
\begin{eqnarray}\label{hydrostatic}
\frac{dp}{ds}(s)=\rho(s) \, g_\parallel(s),
\end{eqnarray}
meaning that the pressure derivative with position must balance the projected gravity force along the field lines. Since we assume an isothermal atmosphere the sound  speed, $c_{\rm s0}$,  is constant and we write
\begin{eqnarray}\label{pressure0}
p(s)=\frac{c_{\rm s0}^2}{\gamma}\rho(s),
\end{eqnarray}
where $\gamma=5/3$. Introducing this expression into Eq.~(\ref{hydrostatic}) we find the following simple differential equation
\begin{eqnarray}\label{hydrostatic1}
\frac{d \rho}{\rho}=\frac{\gamma}{c_{\rm s0}^2}g_\parallel(s) ds,
\end{eqnarray}
which is integrated directly to give 
\begin{eqnarray}\label{dens0}
\rho(s)={\bar{\rho}_0} \exp \left[-\frac{\gamma g}{c_{\rm s0}^2}\frac{L}{\pi} \sin \left(\pi \frac{s}{L}\right)\right],
\end{eqnarray}
where $\bar{\rho}_0$ is the reference density at $s=0$ and $s=L$, that is at the footpoints of the magnetic tube.
According to Eq.~(\ref{pressure0}) gas pressure and density have the same dependence with $s$.

\section{Equilibrium solutions containing threads} \label{equilsect}

A stratified prominence unbounded with height, as for example in the \citet{hoodanzer90} model, can be seen as a continuous collection of threads stacked in the vertical direction but located at magnetic dips. It is crucial to understand the stability of a density enhancement or equivalently a thread in the solar atmosphere, even in the absence of magnetic dips as we explained in Sect.~\ref{intro}.

\begin{figure}[!h]
\center{
\includegraphics[width=9cm]{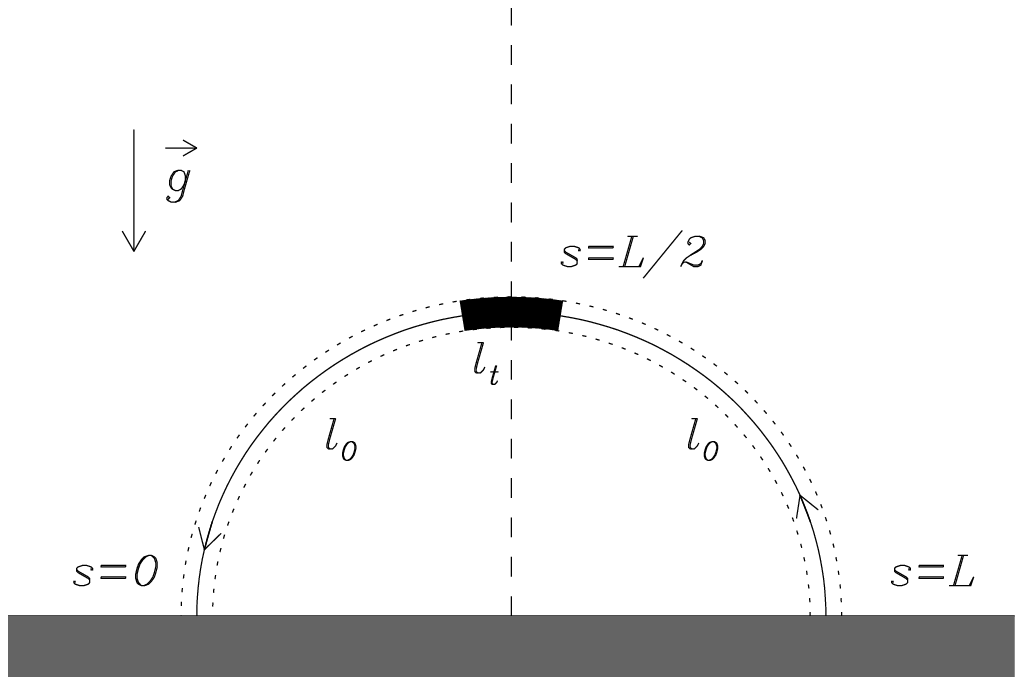}\\
\includegraphics[width=9cm]{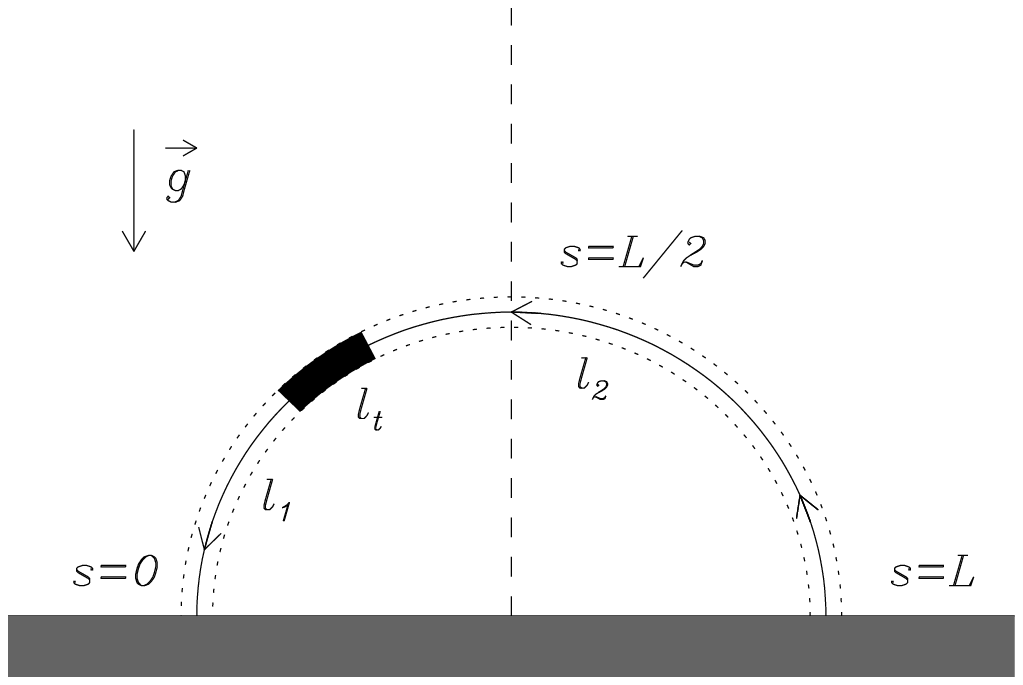}}
\caption{Sketch of the basic model. The upper panel corresponds to the situation in which the plasma thread, represented by the dark rectangle of length $l_{\rm t}$, is located at the tube apex. This configuration is in an equilibrium that can be stable or unstable. The lower panel represents a new equilibrium of the thread at a lower height. The curved magnetic field is uniform and directed along the tube axis. The gravity force is pointing downwards and is projected along the field line according to Eq.~(\ref{gpar}). The footpoints are supposed to be anchored at the base of the corona, represented by the shaded grey area. } \label{figequil}
\end{figure}

The idea is to analyse the simple situation of a finite mass (called a thread, hereafter) with density $\rho_{\rm t}$ located on the curved magnetic field (see Fig.~\ref{figequil} for a sketch of the model). The question is whether this equilibrium is stable or unstable and we investigate the conditions in which the system allows equilibrium solutions at lower heights.

The basic equation of our system is the momentum equation
\begin{eqnarray}\label{momentum}
\rho_{\rm t} \frac{d^2 s}{dt^2}= -\frac{dp}{ds}(s)+\rho_{\rm t}\, g_\parallel(s),
\end{eqnarray}
where $s$ is the coordinate of the centre of the thread along the curved path. Since the thread is surrounded by plasma,  fluctuations in gas pressure at the edges of the thread in the longitudinal direction create a net force. We need to provide an expression for the gas pressure derivative. The other force existing in the system is the  gravity projection along the magnetic field line given in Eq.~(\ref{gpar}). Before we go into the details about the calculation of the gas pressure derivative, it is necessary to describe the general features of our system from a mathematical point of view.

In order to analyse the stability of the system, we write Eq.~(\ref{momentum}) as the vector field
\begin{eqnarray}\label{firstorder}
\left(\begin{array}{c}\Dot{s} \cr \Dot{v}
\end{array}\right)=
\left(\begin{array}{c} v \cr -\frac{1}{\rho_{\rm t} }\frac{dp}{ds}(s)+g_\parallel(s) 
\end{array}\right),
\end{eqnarray}
where the dot represents the time derivative. The system has equilibrium or fixed points when
\begin{eqnarray}\label{moment00}
{v}&=&0,\nonumber\\
-\frac{1}{\rho_{\rm t}}\frac{dp}{ds}(s)+g_\parallel(s)&=& 0,
\end{eqnarray}
and the corresponding solution for the position is denoted as $s_e$; the velocity at the equilibrium point is always zero. 
The linearised version of Eq.~(\ref{firstorder}) is
written as\begin{eqnarray}\label{linsyst0}
\left(\begin{array}{c}\Dot{s} \cr \Dot{v}
\end{array}\right)={\bf A} 
\left(\begin{array}{c} s \cr v
\end{array}\right),
\end{eqnarray}where the matrix {\bf A} is the Jacobian;
and\begin{eqnarray}
{\bf A} = \left(\begin{array}{cc} \frac{\partial \Dot{s}}{\partial s} & \frac{\partial \Dot{s}}{\partial v} \\ [\medskipamount] \frac{\partial \Dot{v}}{\partial s} & \frac{\partial \Dot{v}}{\partial v}
\end{array}\right)=
\left(\begin{array}{cc} 0 & 1 \\ [\medskipamount]  f(s) & 0
\end{array}\right),
\end{eqnarray}
where
\begin{eqnarray}\label{fs}
f(s)=\frac{d}{ds} \left[-\frac{1}{\rho_{\rm t} }\frac{dp}{ds}(s)+g_\parallel(s)\right].
\end{eqnarray}
The eigenvalues of {\bf A} at the equilibrium points ($s=s_{\rm e}$)
provide valuable information about their stability. If $f(s_{\rm e})> 0$,  we have  
\begin{eqnarray}\label{solunstable}
\lambda_\pm = \pm \sqrt{f(s_{\rm e})}.
\end{eqnarray}
In this case the fixed point is a saddle point because we have real values with opposite signs \citep[e.g.][]{jordansmith1987}. This point is unstable.

If $f(s_{\rm e})<0$ , where the linearised system is exactly the same as that of the simple harmonic oscillator equation, the eigenvalues are
written as\begin{eqnarray}\label{solstable}
\lambda_\pm = \pm\, i \, \sqrt{\left|f(s_{\rm e})\right|}.
\end{eqnarray}
This corresponds to a linear centre that is a stable point.
The last case is when $f(s_{\rm e})=0$; this is a degenerate case but still provides information about the stability of the equilibrium points.

\subsection{Thread initially at the tube apex}
\label{sect:equil_sym}

We first analyse the situation of the thread initially located at the top of a curved magnetic field (see Fig.~\ref{figequil} top panel). This mass is in equilibrium because the gravity force is zero at $s=L/2$ (see Eq.~(\ref{gpar})) and the gradient of the background pressure is zero at the tube apex (see Eq.~(\ref{pressure0}) and Eq.~(\ref{dens0})). The goal is to obtain an approximation for the pressure derivative in Eq.~(\ref{moment00}) once the thread has moved from the initial equilibrium position. In our model we apply line-tying conditions, meaning that pressure fluctuations are not lost along the field lines at the base of the corona. There is no energy leakage through the footpoints and mass flows along the magnetic tube injected from the photosphere are not allowed. Therefore we have perfect reflection at these points because of the big difference in density between the photosphere and corona. It is important to remark that this specific boundary condition allows us to have suspended threads at a given height even in the absence of magnetic dips and that the presence of a chromospheric layer (not included in our model) may affect how the pressure is balanced along the tube.

We denote the length of the thread as $l_{\rm t}$ and by a simple geometrical reasoning $l_{\rm t}/2<s<L-l_{\rm t}/2$ because the thread edge cannot connect with the base of the corona. Under equilibrium at $s=L/2$ the length of the external part of tube that contains the thread is $l_0=L/2-l_{\rm t}/2$ along each loop leg (see Fig.~\ref{figequil} top panel). We have the simple relation $L=2l_0+l_{\rm t}$. Now we assume that the thread has moved to another location along the magnetic field (see Fig.~\ref{figequil} bottom panel). The distance on the left side is $l_1$ now (different from $l_0$), while the distance on the right side is $l_2$. We have that $l_1=s-l_{\rm t}/2$, and  $l_2=L-l_1-l_{\rm t}=L-s-l_{\rm t}/2$. Quiescent prominences are typically $100-200\,{\rm Mm}$ long ($L$) and are about a 100-fold denser than the surrounding corona ($\rho_{\rm t}/\rho_0$). However, it is known that the entire bodies of prominences consist in a collection of individual short threads of plasma located in long field lines so that only a fraction of each field line ($l_{\rm t}$) is filled with dense plasma. In this study a wide range of thread parameters are utilised to exemplify a vast sort of prominence threads \citep[see][for a detailed description of solar prominences]{mackayetal10,engvold15}.

\begin{figure}[ht!]
\center
\includegraphics[width=8cm,trim= 0mm 0mm 0mm 0mm]{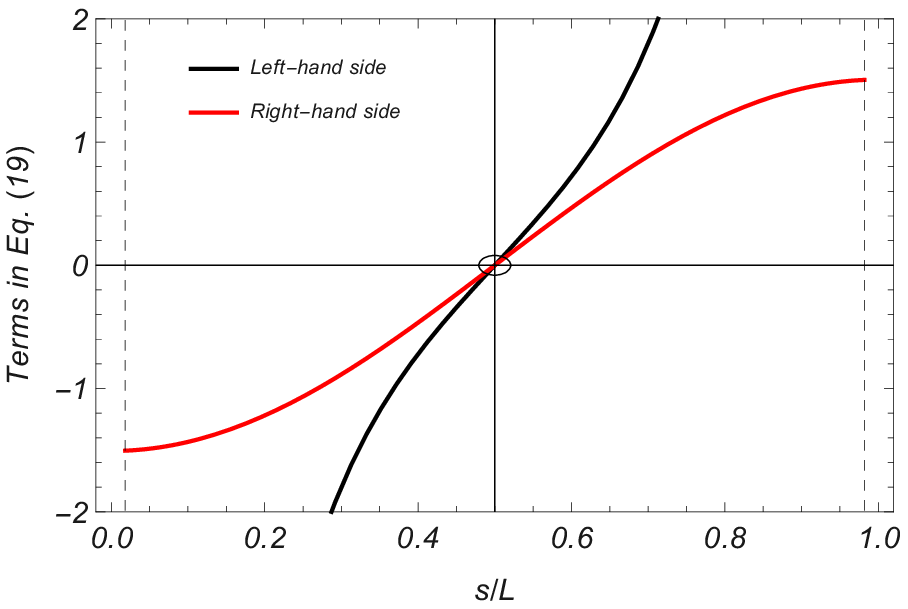}
\includegraphics[width=8cm,trim= 0mm 0mm 0mm 0mm]{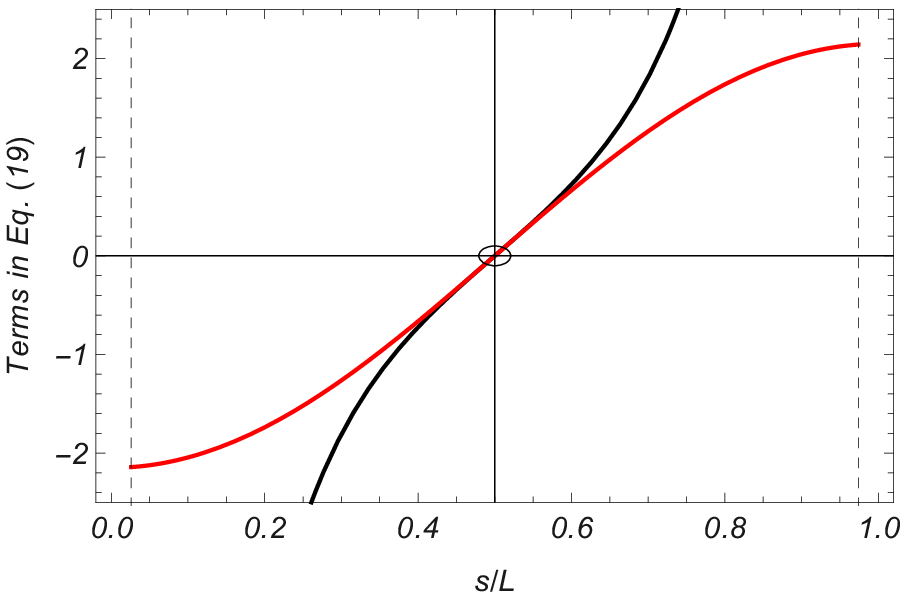}
\includegraphics[width=8cm,trim= 0mm 0mm 0mm 0mm]{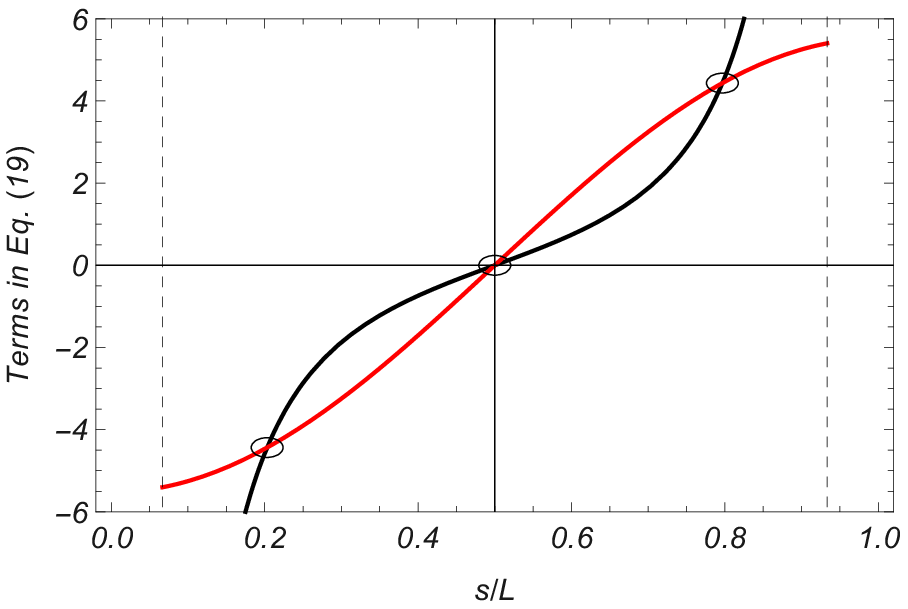}
\caption{Left- (black line) and right- (red line) hand sides of Eq.~(\ref{eqequil}) as a function of $s/L$, for three different values of $l_{\rm t}$. In the top panel the system is linearly stable with one solution (see circle). On the bottom panel the system relaxes to a new equilibrium at a lower height (see the two complementary solutions) and the trivial solution becomes an unstable solution. In all cases, $L=10\,H$ and $\rho_{\rm t}=100~\rho_0$; $l_{\rm t}=L/110$ (top panel), $l_{\rm t}={l_{\rm t}}_{\rm b}$ (middle panel), and  $l_{\rm t}=L/30$ (bottom panel). The value {\bf ${l_{\rm t}}_{\rm b}\approx L/77.08$} is obtained when the derivatives of the two curves at $s=L/2$ are the same (see Eq.~(\ref{eqequilbif})). The vertical dashed lines represent the edges of the domain of the variable $s$, namely $s=l_{\rm t}/2$ and $s=L-l_{\rm t}/2$.} \label{figsols}
\end{figure}

Since the system is assumed to be dissipationless, we  consider that the evolution during the motion of the thread is adiabatic. The behaviour of the plasma in the evacuated parts of the tube, that is where the density is low, is similar to that of a piston because of the line-tying conditions at the footpoints.
The gas pressure can be written as the background, stratified from gravity, plus a term that depends on the position of the thread. The key idea is to apply the adiabatic law to each part of the tube, but to this term only, removing the contribution from the background,
\begin{eqnarray}\label{adiabatic}
p_1 \left(l_1 A_1\right)^\gamma=p_0 \left(l_0 A_0\right)^\gamma,\nonumber\\ 
p_2 \left(l_2 A_2\right)^\gamma=p_0 \left(l_0 A_0\right)^\gamma,
\end{eqnarray}
where $p_0$ represents the reference pressure before the thread has moved, changing the length of the evacuated part from $l_0$ to $l_1$ and pressure from $p_0$ to $p_1$ for the left side. For the right side we have changes from $l_0$ to $l_2$ and from $p_0$ to $p_2$. Since we are considering a perfectly circular geometry we have that $A_0=A_1=A_2$ and the cross-sectional areas cancel in the previous expressions. Now from Eq.~(\ref{adiabatic}) we obtain that
\begin{eqnarray}\label{adiabatic1}
p_1(s)=p_0 \left(\frac{l_0}{l_1}\right)^\gamma=p_0 \left(\frac{L/2-l_{\rm t}/2}{s-l_{\rm t}/2}\right)^\gamma,\nonumber\\ 
p_2(s)=p_0 \left(\frac{l_0}{l_2}\right)^\gamma=p_0 \left(\frac{L/2-l_{\rm t}/2}{L-s-l_{\rm t}/2}\right)^\gamma.
\end{eqnarray}

\noindent At this point we impose the condition that the plasma thread has to be in static and stationary equilibrium, namely, no flows are allowed and the solution does not change in time. These conditions imply that Eq.~(\ref{firstorder}) reduces to Eq.~(\ref{moment00}). To simplify things we assume that the thread does not change its shape and keeps its length, $l_{\rm t}$, constant. Now we approximate the gas pressure derivative that appears in Eq.~(\ref{moment00}) as  
\begin{eqnarray}\label{dpdsaprox}
\frac{dp}{ds}(s) \approx \frac{\Delta p}{\Delta s}(s)=\frac{p_2(s)-p_1(s)}{l_{\rm t}}.
\end{eqnarray}
Since the thread has a finite length, gravity changes along this thread; therefore this term is approximated by the average gravity on the thread
\begin{eqnarray}\label{gparav}
{\bar g}_\parallel(s)&=&\frac{1}{l_{\rm t}}\int^{s+l_{\rm t}/2}_{s-l_{\rm t}/2} g_\parallel(s)ds\nonumber \\ &=&-g\,\frac{2}{\pi}\, \frac{L}{l_{\rm t}} \sin\left(\frac{\pi}{2} \frac{l_{\rm t}}{ L}\right)\cos\left(\pi \frac{s}{L}\right).
\end{eqnarray}
Since under typical conditions $l_{\rm t}\ll L$, using the Maclaurin expansion of the sinus for small arguments, we recover exactly $g_\parallel(s)$ in the previous expression. Hereafter, to simplify things we use $g_\parallel(s)$ from Eq.~(\ref{gpar})  instead of Eq.~(\ref{gparav}).

The hydrostatic equation for the thread is therefore
\begin{eqnarray}\label{aproxmoment1}
\frac{p_2(s)-p_1(s)}{l_{\rm t}}=\rho_{\rm t}\,{g}_\parallel(s). 
\end{eqnarray}
Using Eq.~(\ref{pressure0}) for the sound speed together with the previous expressions for pressure and gravity we obtain the following equation for $s$, representing the location under static equilibrium  of the dense thread along the tube,
\begin{eqnarray}\label{eqequil}
\left(\frac{L/2-l_{\rm t}/2}{L-s-l_{\rm t}/2}\right)^\gamma&-&\left(\frac{L/2-l_{\rm t}/2}{s-l_{\rm t}/2}\right)^\gamma=\nonumber\\&-&\frac{\gamma}{c^2_{{\rm s0}}} \frac{\rho_{\rm t}}{\rho_0}\, g \, l_{\rm t} \,\cos\left(\pi \frac{s}{L}\right),
\end{eqnarray}
where $c^2_{{\rm s0}}=\gamma p_0/\rho_0$, and $\rho_{\rm t}/\rho_0$ corresponds to the density contrast between the thread and the density background at $s=L/2-l_{\rm t}/2$. The parameter $\rho_{\rm t}/\rho_0$ must be much larger than 1 since we are not taking into account any energy flux through the thread and its mass and length are assumed to be constant. 

Equation~(\ref{eqequil}) is transcendental and has to be solved numerically. From now on we denote $s_{\rm e}$ a solution of Eq.~(\ref{eqequil}). It is easy to verify that $s_{\rm e}=L/2$ is a solution for any values of the parameters.  Hence, there is always an equilibrium solution at the tube apex ($s_{\rm e}^\mathscr{A}= L/2$) that can be stable or unstable. Furthermore, the symmetry of the model with respect to $s=L/2$ implies that if there are additional solutions, they appear in pairs ($s_{\rm e}^{\mathscr{L}}, s_{\rm e}^{\mathscr{R}}$) verifying $s_{\rm e}^\mathscr{L}+s_{\rm e}^\mathscr{R}=L$, to wit, the solutions are distributed symmetrically with respect to $s=L/2$: one on the left ($s_{\rm e}^\mathscr{L}$) and the other on the right side ($s_{\rm e}^\mathscr{R}$) of the tube.

To understand the nature of the these possible situations, in Fig. 2 we plotted the left- and right-hand side terms of Eq.~(\ref{eqequil}) as a function of $s$, for fixed parameters $L$, $c_{{\rm s0}}$, $g$, $\rho_{\rm t}/\rho_0$, and varying $l_{\rm t}$ using the projected gravity given in Eq.~(\ref{gpar}). The gravity acceleration is $g=274\,{\rm m\,s^{-2}}$, and we take a spatial reference length of $H=10,000\,{\rm km}$. The sound speed takes a value of $c_{{\rm s0}}=166\,{\rm km}\,{\rm s^{-1}}$ for a fully ionised coronal tube with $10^6$~K temperature. When the two curves cross  (see circles) we have a solution at that particular point. On the top panel we have only one solution: the trivial one at $s_{\rm e}^\mathscr{A}=L/2$. On the bottom panel however three solutions appear: the trivial one plus two complementary solutions. As we describe in Sect.~\ref{wavessect}, in the first situation the mass remains at its original position, and, upon being subject to a small disturbance, it oscillates around this point because this equilibrium is stable. In the second situation, the mass is not stable at the initial position at the apex, and after being perturbed the mass moves to one of the solutions along the legs that are energetically more favourable.  The stability analysis performed in this  and in the following sections provides a simple criterion that tells us when a plasma thread is gravitationally stable or unstable.

The transition between one to three solutions is represented in the middle panel of Fig.~\ref{figsols} and the mathematical condition is that the slopes of the curves are the same for the left- and right-hand sides of  Eq.~(\ref{eqequil}),
\begin{eqnarray}\label{eqequilbif}
\frac{\left(L/2-l_{\rm t}/2\right)^\gamma}{(L-s-l_{\rm t}/2)^{\gamma+1}}&+&\frac{\left(L/2-l_{\rm t}/2\right)^\gamma}{(s-l_{\rm t}/2)^{\gamma+1}}=\nonumber \\ && \frac{1}{c^2_{{\rm s0}}} \frac{ \rho_{\rm t}}{\rho_0}\,g\, l_{\rm t}\, \frac{\pi}{L}\sin\left(\pi \frac{s}{L}\right).
\end{eqnarray}
\noindent
Considering the following equation:
\begin{eqnarray}\label{dsterm}
f(s)=\frac{d}{ds} \left[-\frac{1}{\rho_{\rm t} }\frac{dp}{ds}(s)+g_\parallel(s)\right]=0,
\end{eqnarray}
and using the expressions for the pressure derivative (Eq.~(\ref{dpdsaprox})) and the projected gravity (Eq.~(\ref{gpar})), we obtain Eq.~(\ref{eqequilbif}). When Eq.~(\ref{eqequilbif}) or Eq.~(\ref{dsterm}) is satisfied together with Eq.~(\ref{eqequil}) the system is said to have a bifurcation point. We realise that the left-hand term of Eq.~(\ref{dsterm}) is precisely the term inside the square root of the computed eigenvalues in Eqs.~(\ref{solunstable}) and (\ref{solstable}). We have shown that depending on the sign of $f(s_{\rm e})$ the nature of the equilibrium solutions changes. When $f(s_{\rm e})$ is positive (Eq.~(\ref{solunstable})) the eigenvalue $\lambda_-$  represents an exponentially damped solution that is eventually dominated by the growing solution in time associated to $\lambda_+$ of the form $e^{\tau t}$  where the growth time is $\tau=\lambda_+$.  When $f(s_{\rm e})$ is negative we have a stable linear centre with a harmonic time-dependence of the form $e^{i \omega t}$ and frequency $\omega=\lambda_+/i$ (Eq.~(\ref{solstable})).
In this case, it is known from linear theory of ODE systems that we cannot conclude that the equilibrium point is truly a centre since it could be also a stable or unstable spiral \citep[e.g.][]{jordansmith1987}. Nevertheless, when the system is reversible or energy conservative (Hamiltonian), and in our case the two conditions are fulfilled, the 
linear centre is a true centre \citep[e.g.][Sections 6.5 and 6.6]{strogatz2018} and stability is ensured. In Sect.~\ref{wavessect}  we confirm, using physical arguments, that the equilibrium point is stable.

\begin{figure}[ht!]
        \begin{center}
        \includegraphics[trim = 0mm 0mm 0mm 0mm,clip, width=8.cm]{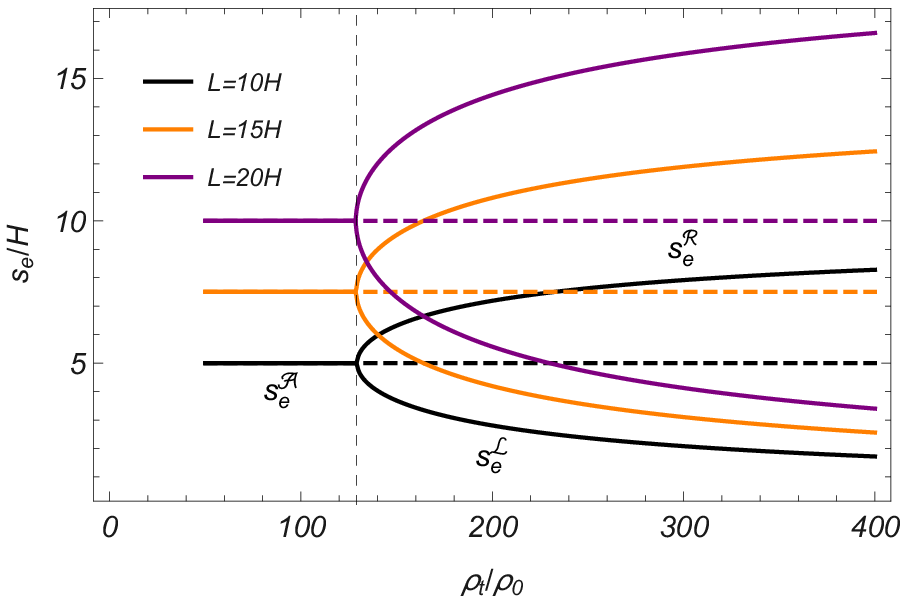}
        \includegraphics[trim = 0mm 0mm 0mm 0mm,clip, width=8.cm]{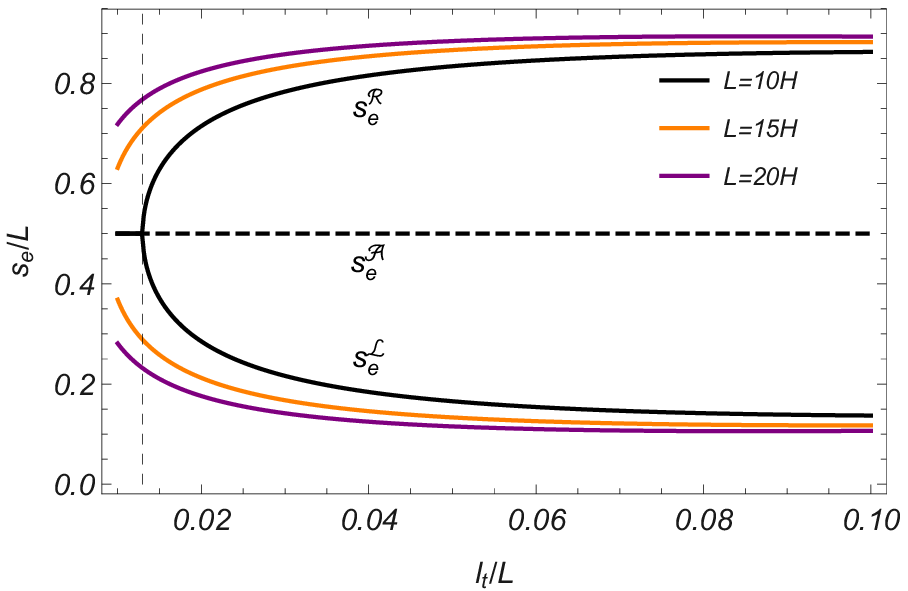}
        \includegraphics[trim = 0mm 0mm 0mm 0mm,clip, width=8.cm]{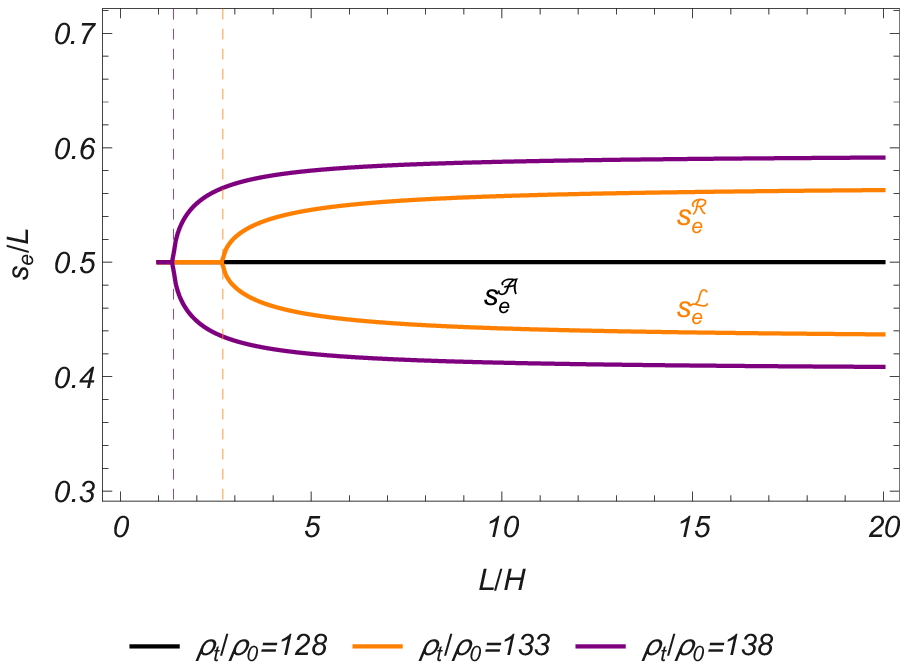}
        \caption{Solutions $s_{\rm e}$ of Eq.~(\ref{eqequil}) as a function of $\rho_{\rm t}/\rho_0$ (top panel), $l_{\rm t}/L$ (middle panel), and $L/H$ (bottom panel). The expression $l_{\rm t}=0.1\,H$ in the top and bottom panels and $\rho_{\rm t}/\rho_0=100$ in the middle panel. The solid lines correspond to stable solutions and dashed lines to unstable solutions. The thin vertical dashed lines correspond to the bifurcation values that have been calculated by solving Eqs.~(\ref{eqequil}) and (\ref{eqequilbif}) simultaneously and coincide with the results in  Eq.~(\ref{rhoc}) (top panel), Eq.~(\ref{l_t}) (middle panel), and Eq.~(\ref{lc}) (bottom panel). }
        \label{fig:equilbifurc}
\end{center}
\end{figure}
We discuss the results from the physical point of view. Figure~\ref{figsols} shows that the existence of the non-trivial solutions of Eq.~(\ref{eqequil}) depends on the thread model parameters. Figure~\ref{fig:equilbifurc} (top panel) shows the equilibrium position $s_{\rm e}$ as a function of $\rho_{\rm t}/\rho_0$ for different values of $L$ when the thread length is $l_{\rm t}=0.1\,H$. We obtain that the system only develops the $s_{\rm e}^\mathscr{L}$ and the $s_{\rm e}^\mathscr{R}$ solutions for values of $\rho_{\rm t}$ beyond ${\rho_{\rm t}}_{\rm b}$. When $\rho_{\rm t}<{\rho_{\rm t}}_{\rm b}$, meaning before the appearance of the bifurcation, $s_{\rm e}^\mathscr{A}$ is the only equilibrium position, so that we are in the situation of the top panel of Fig.~\ref{figsols} and therefore $s_{\rm e}^\mathscr{A}/L=0.5$ is stable. However, when the thread structure enables the existence of two complementary solutions, these are energetically more suitable so that $s_{\rm e}^\mathscr{A}$, which remains located at  $s_{\rm e}/L=0.5$ (see horizontal dashed lines in Fig.~\ref{fig:equilbifurc} top panel), becomes an unstable solution. To distinguish between stable or unstable solutions, we represent in Fig.~\ref{fig:equilbifurc} (and in the following figures) stable solutions as solid curves and unstable solutions as dashed curves. In addition, we realise that for the three values of $L$, $s_{\rm e}^\mathscr{L}$ and $s_{\rm e}^\mathscr{R}$ appear around the same bifurcation value ${\rho_{\rm t}}_{\rm b}$. This means that the bifurcation density value from which we obtain an equilibrium out of the apex of the structure, does not strongly depend on the length of the magnetic tube. We note that $s_{\rm e}$ is normalised to $H$ for normalisation purposes since if we keep the normalisation to $L$, the solutions of the three configurations overlap. This means that the normalised $s_{\rm e}/L$ does not depend strongly on $L$ for this set of parameters. This feature is clearer in the bottom panel of Fig.~\ref{fig:equilbifurc}, where $s_{\rm e}$ as a function of $L$ is plotted; when $L$ is large the stable solutions are almost horizontal. Moreover, as was expected, the equilibrium position of the thread drops with increasing $\rho_{\rm t}$. It is interesting to study the dependence of the equilibrium position on the thread length. Figure~\ref{fig:equilbifurc} (middle panel) shows $s_{\rm e}$ as a function of $l_{\rm t}/L$ for different values of $L$ when $\rho_{\rm t}/\rho_0=100$. For $L=10\,H$ we obtain that below the bifurcation $s_{\rm e}=s_{\rm e}^\mathscr{A}$ is the only solution of Eq.~(\ref{eqequil}) and this is stable, but when $l_{\rm t}>{l_{\rm t}}_{\rm b}$, $s_{\rm e}^\mathscr{A}$ becomes unstable and the emerged branches $s_{\rm e}^\mathscr{L}$ and $s_{\rm e}^\mathscr{R}$ are stable. For $L=15\,H$ and $L=20\,H$ the bifurcation thread length is out of the plot and their $s_{\rm e}^\mathscr{A}$ overlaps with that for $L=10\,H$ at $s_{\rm e}/L=0.5$. In Fig.~\ref{fig:equilbifurc} (middle panel) we also see that the equilibrium position of the thread changes along the magnetic tube when the thread length increases. Figure \ref{fig:equilbifurc} (top and middle panels) show the same behaviour for all sets of parameters, which consists in a sole stable solution for values below the bifurcation, and two symmetric stable solutions out of the apex plus an unstable solution at $s_{\rm e}=L/2$ beyond the bifurcation point. However, Fig.~\ref{fig:equilbifurc} (bottom panel) shows that for some set of parameters, the system does not present the $s_{\rm e}^\mathscr{L}$ and $s_{\rm e}^\mathscr{R}$ equilibria when $L$ varies so that $s_{\rm e}^\mathscr{A}$ is always stable. Interestingly, the bottom panel of Fig.~\ref{fig:equilbifurc} shows that for small changes of $\rho_{\rm t}$, $s_{\rm e}^\mathscr{L}$ and  $s_{\rm e}^\mathscr{R}$ vary significantly. Moreover, we see that the bifurcation length $L_{\rm b}$ for $\rho_{\rm t}/\rho_0=133$ and $\rho_{\rm t}/\rho_0=138$ is short, around $L_{\rm b}=2.7\,H$ and $L_{\rm b}=1.4\,H$, respectively. These values are shorter than the typical magnetic tube observations. 

More aspects about the bifurcation values and how to calculate them are given in Sect.~\ref{wavessect}. But we finish this section by analysing the presented results from a dynamical systems perspective. The transition from one stable equilibrium solution to two symmetric stable solutions and one unstable solution is characteristic of the supercritical pitchfork bifurcation (see e.g. Sect. 3.4 of \citet{strogatz2018} or Sect. 20.1{\rm E} of \citet{wiggins2003}). This bifurcation is present in systems that have a symmetry, such as the symmetry between the left and right sides of our magnetic tube when the thread is initially at the tube apex. In Fig.~\ref{fig:equilbifurc} we find the same type of pitchfork bifurcation in the three panels, although in the bottom panel the unstable solutions for $\rho_{\rm t}/\rho_0=133$ and $\rho_{\rm t}/\rho_0=138$ are overplotted by the stable solution of $\rho_{\rm t}/\rho_0=129$.

\begin{figure}[ht!]
\center
\includegraphics[width=8cm,trim= 0mm 0mm 0mm 0mm]{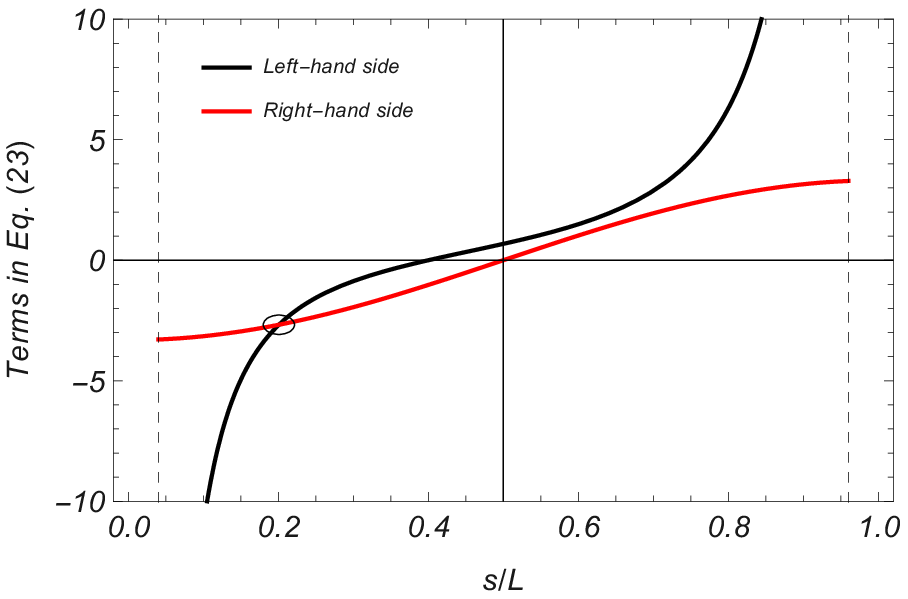}
\includegraphics[width=8cm,trim= 0mm 0mm 0mm 0mm]{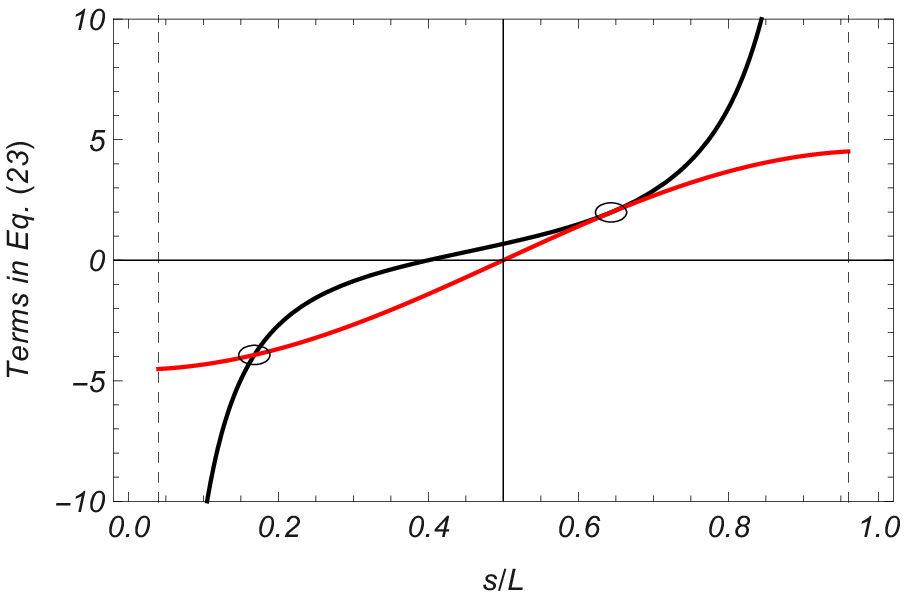}
\includegraphics[width=8cm,trim= 0mm 0mm 0mm 0mm]{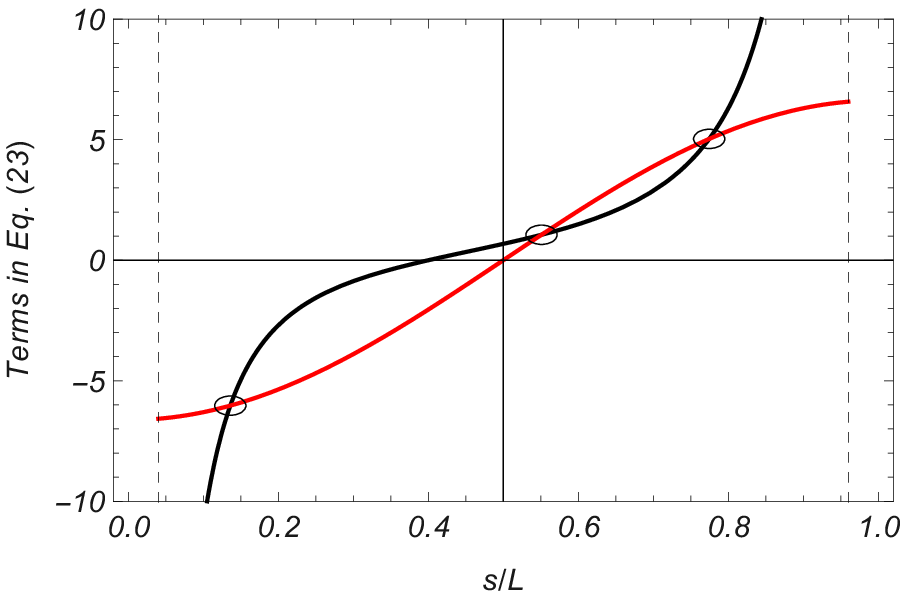}
\caption{Left- (black line) and right- (red line) hand sides of Eq.~(\ref{eqequil2}) as a function of $s/L$, for three different situations with one, two, and three solutions. In these plots $s_0=0.4/L$ (where the black line crosses zero), $L=10\,H$, $l_{\rm t}=L/50$; $\rho_{\rm t}/\rho_0=100$ (top panel), $\rho_{\rm t}/\rho_0={\rho_{\rm t}}_{\rm b}/\rho_0$ (middle panel), and $\rho_{\rm t}/\rho_0=200$ (bottom panel). The value ${\rho_{\rm t}}_{\rm b}/\rho_0\approx137.28$ is obtained from Eqs.~(\ref{eqequil2}) and (\ref{eqequil2d}). The vertical dashed lines represent the edges of the domain of the variable $s$, namely $s=l_{\rm t}/2$ and $s=L-l_{\rm t}/2$.} \label{figsols2}
\end{figure}

\begin{figure}[ht!]
        \begin{center}
        \includegraphics[trim = 0mm 0mm 0mm 0mm,clip, width=8.cm]{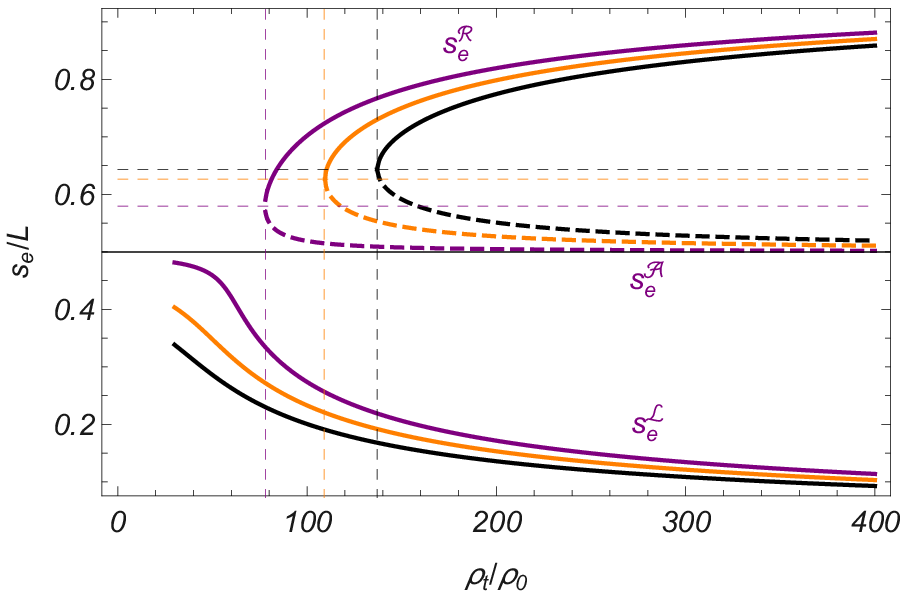}
        \includegraphics[trim = 0mm 0mm 0mm 0mm,clip, width=8.cm]{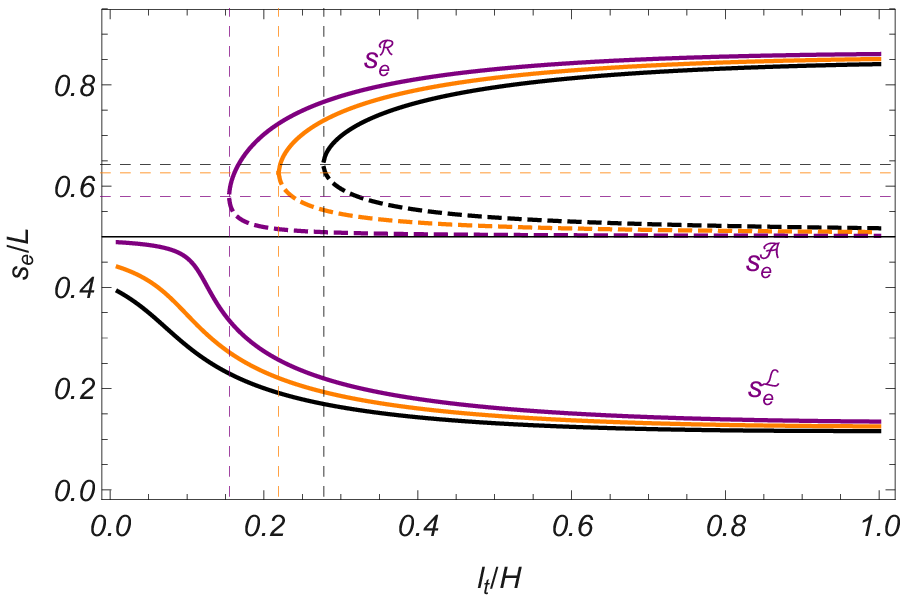}
        \includegraphics[trim = 0mm 0mm 0mm 0mm,clip, width=8.cm]{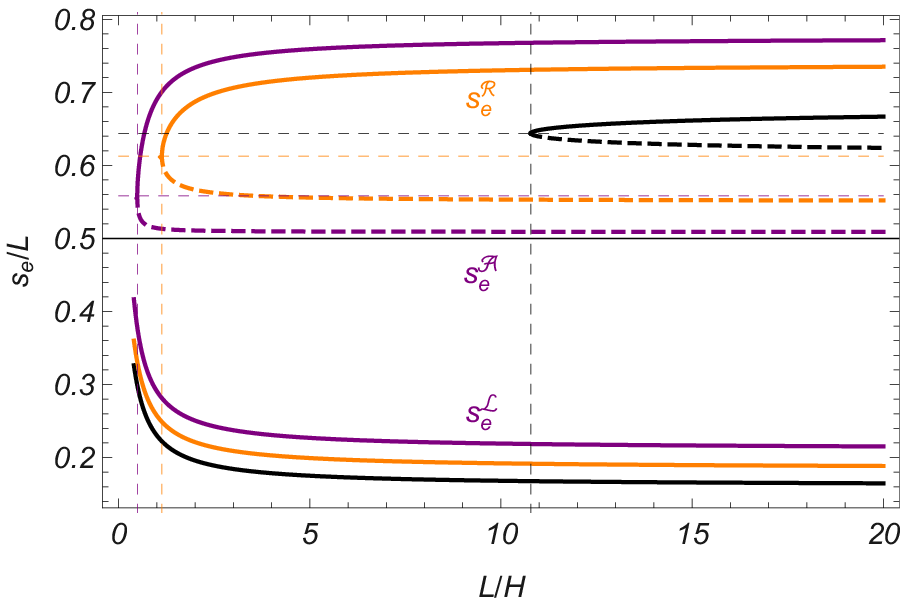}
        \includegraphics[trim = 0mm 0mm 0mm 0mm,clip, width=6.cm]{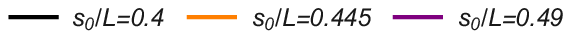}
        \caption{Same as Fig.~\ref{fig:equilbifurc} but for Eq.~(\ref{eqequil2}) and for different values of $s_0/L$. $L=10\,H$ and $l_{\rm t}=0.2\,H$ (top panel); $\rho_{\rm t}/\rho_0=100$ and $L=10\,H$ (middle panel); $l_{\rm t}=0.2\,H$ and $\rho_{\rm t}/\rho_0=137$ (bottom panel). The bifurcation points (intersections between vertical and horizontal thin dashed lines) have been calculated by solving Eqs.~(\ref{eqequil2}) and (\ref{eqequil2d}) simultaneously.}
        \label{fig:equilbifurcnosim}
\end{center}
\end{figure}

\subsection{Thread initially at any location along the tube}\label{nonsymmetric}

It is straightforward to derive analytical expressions for the situation of a thread initially located at any position along the tube, denoted by $s_0$, and not necessarily at $L/2$ as discussed in the previous section. We simply need to rewrite Eq.~(\ref{adiabatic1}) as
\begin{eqnarray}\label{adiabatic2}
p_1(s)&=&p_0 \left(\frac{s_0-l_{\rm t}/2}{s-l_{\rm t}/2}\right)^\gamma,\nonumber\\ 
p_2(s)&=&p_0 \left(\frac{L-s_0-l_{\rm t}/2}{L-s-l_{\rm t}/2}\right)^\gamma,
\end{eqnarray}
and the corresponding equilibrium equation analogous to Eq.~(\ref{eqequil}) becomes
\begin{eqnarray}\label{eqequil2}
\left(\frac{L-s_0-l_{\rm t}/2}{L-s-l_{\rm t}/2}\right)^\gamma&-&\left(\frac{s_0-l_{\rm t}/2}{s-l_{\rm t}/2}\right)^\gamma=\nonumber \\ &-&\frac{\gamma}{c^2_{{\rm s0}}} \frac{\rho_{\rm t}}{\rho_0}\, g \, l_{\rm t} \,\cos\left(\pi \frac{s}{L}\right).
\end{eqnarray}
Equation (\ref{eqequil2}) simplifies to Eq.~(\ref{eqequil}) when $s_0=L/2$. Different types of representative solutions are shown in Fig.~\ref{figsols2} and in this case the parameter that has been chosen to change is $\rho_{\rm t}$ (in Fig.~\ref{figsols} it was $l_{\rm t}$). In these plots the initial position of the thread is $s_0=0.4/L$, therefore located on the left half of the tube. The left-hand side of Eq.~(\ref{eqequil2}) is not symmetric anymore with respect to the tube apex, and this facilitates the existence either of one, two, or three solutions, according to Fig.~\ref{figsols2}. But we note that by continuity there must be a situation in these circumstances with two solutions when the left- and right-hand curves are tangential. This case is calculated as in  Sect.~\ref{sect:equil_sym} by imposing that the spatial derivatives of the left- and right-hand sides of  Eq.~(\ref{eqequil2}) are equal as follows:
\begin{eqnarray}\label{eqequil2d}
\frac{\left(L-s_0-l_{\rm t}/2\right)^\gamma}{(L-s-l_{\rm t}/2)^{\gamma+1}}&+&\frac{\left(s_0-l_{\rm t}/2\right)^\gamma}{(s-l_{\rm t}/2)^{\gamma+1}}=\nonumber \\& &\frac{1}{c^2_{{\rm s0}}} \frac{\rho_{\rm t}}{\rho_0}\, g \, l_{\rm t}\frac{\pi}{L} \,\sin\left(\pi \frac{s}{L}\right).
\end{eqnarray}
Thus, Eq.~(\ref{eqequil2d}) has to be satisfied together with Eq.~(\ref{eqequil2}) for the system to have two equilibrium solutions. An example of such a situation is found in Fig.~\ref{figsols2} middle panel. For simplicity, we still denote the equilibrium position on the left side of the tube as $s_{\rm e}^\mathscr{L}$, as $s_{\rm e}^\mathscr{R}$ on the right side, and as $s_{\rm e}^\mathscr{A}$ for the central near apex solution, so that when the system has only two equilibrium solutions, $s_{\rm e}^\mathscr{R}=s_{\rm e}^\mathscr{A}$ (or $s_{\rm e}^\mathscr{L}=s_{\rm e}^\mathscr{A}$). As in Sect.~\ref{sect:equil_sym} the sign of the difference between the left-hand term and the right-hand term of Eq.~(\ref{eqequil2d}) reveals the nature of the different solutions.

In the bifurcation diagram we find that the characteristic pitchfork for the symmetric case disconnects into two curves: one below $s_0/L=0.5$ and another above this value. An example is shown in  Fig.~\ref{fig:equilbifurcnosim} (top panel) when the density contrast is changed.  The lower curve  is always stable, whereas the upper curve has both stable and unstable branches. Therefore, the pitchfork bifurcation occurring in the symmetrical case is essentially replaced in the non-symmetrical case by a stable state (lower curve) alongside a saddle-node bifurcation (upper curve). As we increase the density contrast there is no longer a sharp transition at the bifurcation point representative of the symmetric case (compare with Fig.~\ref{fig:equilbifurc}). Hence, the equilibrium position of the thread varies smoothly with the density contrast on the lower stable curve. Furthermore, the stable branch of the upper curve, which is also a permitted equilibrium state of the system, is not accessible unless the thread is given a large initial disturbance, thereby making this equilibrium more favourable. Qualitatively similar results are obtained for the bifurcation diagrams as a function of the thread length and the total length in the middle and bottom panels of Fig. 5, respectively. As was expected, when $s_0/L\rightarrow0.5$, the diagrams approach those in Fig.~\ref{fig:equilbifurc}, characterised by a stable solution at $s_{\rm e}/L=0.5$ below the bifurcation values; however we see that the two curves (upper and lower) still remain disconnected for $s_0/L=0.49$.

\begin{figure}[ht!]
        \begin{center}
        \includegraphics[trim = 0mm 0mm 0mm 0mm,clip, width=8.cm]{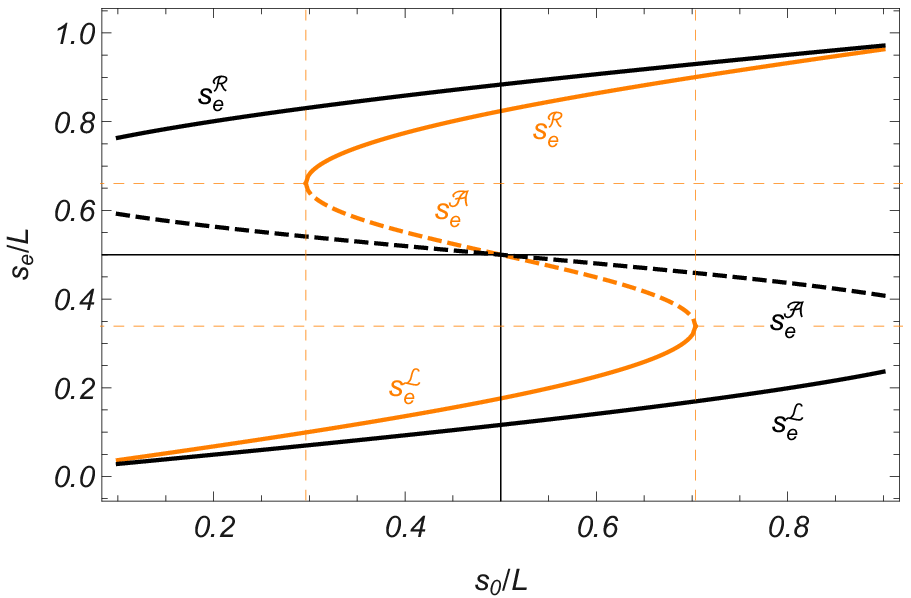}\\
        \includegraphics[trim = 0mm 0mm 0mm 0mm,clip, width=4.cm]{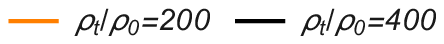}
        \caption{Solutions $s_{\rm e}$ of Eq.~(\ref{eqequil2}) as a function of $s_0/L$ for $\rho_{\rm t}/\rho_0=200$ (orange lines) and $\rho_{\rm t}/\rho_0=400$ (black lines). The expression $L=10\,H$ and $l_{\rm t}=0.2\,H$. The curves correspond to $s_{\rm e}^\mathscr{L}$  and $s_{\rm e}^\mathscr{R}$ (solid lines) and  $s_{\rm e}^\mathscr{A}$ (dashed line). The bifurcation points around $s_0/L=0.3$ and $s_0/L=0.7$ (intersections between vertical and horizontal orange dashed lines) have been calculated by solving Eqs.~(\ref{eqequil2}) and (\ref{eqequil2d}) simultaneously.}
        \label{fig:se_s0_two}
\end{center}
\end{figure}

Figure~\ref{fig:se_s0_two} shows the situation when there is a transition between one to three solutions (orange lines, $\rho_{\rm t}/\rho_0=200$) when the initial position of the thread, $s_0$, is changed. One of the emerging solutions around $s_0/L=0.3$, $s_{\rm e}^\mathscr{A}$,  is unstable, while the other solution, $s_{\rm e}^\mathscr{R}$, is stable. This behaviour is similar to that found for the symmetric case with $s_0=L/2$ described earlier. In Fig.~\ref{fig:se_s0_two} a case with no bifurcations but with multiple solutions for fixed parameter values ($\rho_{\rm t}/\rho_0=400$) is also represented. The solutions that are around the apex (dashed curves) have a different behaviour with respect to the left- and right-hand solutions. In this case $s_{\rm e}^\mathscr{A}$ decreases when $s_0$ is raised, while for the other two solutions $s_{\rm e}^{\mathscr{L}}$ and  $s_{\rm e}^{\mathscr{R}}$ increase with $s_0$. When $s_0$ moves closer to the tube apex, $s_{\rm e}^\mathscr{A}\rightarrow L/2$ and this solution is unstable. In conclusion, given that the initial thread position, $s_0$, is not at the apex, this parameter breaks the symmetry of the system. In bifurcation theory it is called an imperfection parameter. The resulting bifurcation diagram of Fig.~\ref{fig:se_s0_two} is no longer symmetric about a horizontal line and possesses an S-shape (see \citet{wang1994} and Sect.~3.6 of \citet{strogatz2018}). The orange curve in this figure has a lower and an upper bifurcation point, $s_{0,{\rm l}}/L\simeq 0.3$ and $s_{0,{\rm u}}/L\simeq 0.7$, respectively. This S-shaped bifurcation curve can be thought of as the composition of two saddle-node bifurcation curves, the lower (upper) one with its bifurcation point at $s_0=s_{0,{\rm l}}$ ($s_0=s_{0,{\rm u}}$). The black curve of Fig.~\ref{fig:se_s0_two} is a special case of an S-shaped bifurcation curve because the  lower and upper bifurcation points are outside the range of validity of the parameter $s_0$. Therefore, the system always supports three solutions.

\section{Linear stability analysis}
\label{wavessect}

We have intuitively described the nature of  stable and unstable solutions in the previous section and complemented this description with known results from bifurcation theory in mathematics. In the present section we provide a precise stability analysis based on linear theory but using physical grounds. We derive an expression for the oscillatory frequency of the thread, $\omega$, as a function of the equilibrium parameters. The sign of $\omega^2$ renders estimable knowledge about the stability properties of the configuration.

\subsection{Thread initially at the tube apex}\label{analyticexpr}

We start with the trivial solution at the tube apex. We consider a small longitudinal displacement, $\delta s$, of the centre of the plasma thread. We assume that $\delta s\ll L$ because we are interested in the linear regime. The restoring forces acting on the plasma thread are the pressure gradient and the projected gravity force. Let us write the pressure at both sides of the thread as a function of the displacement with respect to $L/2$, the equilibrium position. Using the adiabatic assumption again, the pressure on the left and right parts of the tube is
given by\begin{eqnarray}\label{adiabaticwave}
p_1(\delta s)&=&p_0 \left(\frac{L/2-l_{\rm t}/2}{L/2-l_{\rm t}/2+\delta s}\right)^\gamma,\nonumber\\ 
p_2(\delta s)&=&p_0 \left(\frac{L/2-l_{\rm t}/2}{L/2-l_{\rm t}/2-\delta s}\right)^\gamma.
\end{eqnarray}
The terms in parenthesis are written as
\begin{eqnarray}\label{adiabaticwave1}
\left(\frac{L/2-l_{\rm t}/2}{L/2-l_{\rm t}/2\pm\delta s}\right)^\gamma&=&\left(\frac{1}{1\pm2\delta s/(L-l_{\rm t})}\right)^\gamma
\nonumber \\ &=&\left({1\pm2\delta s/(L-l_{\rm t})}\right)^{-\gamma}\nonumber \\&\simeq& 1\mp2 \gamma \delta s/(L-l_{\rm t}),
\end{eqnarray}
where in the last step we used the Maclaurin series approximation for small arguments (because $\delta s/L\ll 1$). This approximation is crucial to find the linear result we are looking for, since pressure fluctuations are now linearly proportional to $\delta s$. We have  that  
\begin{eqnarray}\label{adiabaticwaveapprox}
p_1( \delta s)&\simeq &p_0 \left[1-2 \gamma\, \delta s/(L-l_{\rm t})\right],\nonumber\\ 
p_2(\delta s)&\simeq&p_0 \left[1+2 \gamma\, \delta s/(L-l_{\rm t})\right],
\end{eqnarray}
and the pressure gradient is given by
\begin{eqnarray}\label{aproxpresswave}
\frac{dp}{ds}(\delta s) \simeq \frac{\Delta p}{\Delta s}(\delta s)=\frac{p_2-p_1}{l_{\rm t}}=\frac{4\, p_0\, \gamma\, \delta s}{l_{\rm t}\, (L-l_{\rm t})}.
\end{eqnarray}
For the projected gravity around $s=L/2$ it is not difficult to see that Eq.~(\ref{gpar}) reduces to a sinus of $\delta s$,
\begin{eqnarray}\label{gpar1}
g_\parallel(L/2+\delta s)=g\, \sin \left(\pi \frac{\delta s}{L}\right) \simeq g \frac{\pi}{L} \delta s,
\end{eqnarray}
where we again used the approximation for small arguments in the last step.

We have expressions for all the terms that appear in the momentum equation, that is Eq.~(\ref{momentum}) for $s=L/2+\delta s$, which is now  
\begin{eqnarray}\label{displ}
\rho_{\rm t} \frac{d^2\delta s}{dt^2}&=& -\frac{4\, p_0\, \gamma }{l_{\rm t}\, (L-l_{\rm t})} \delta s+\rho_{\rm t}\, g\, \frac{\pi}{L} \delta s\nonumber \\ &=&\delta s \left[ -\frac{4\, p_0\, \gamma }{l_{\rm t} (L-l_{\rm t})} +\rho_{\rm t}\, g\, \frac{\pi}{L}\right].
\end{eqnarray}
The right-hand side term of this equation is proportional to $\delta s$ only. The solution to the differential equation is of the form $e^{i \omega t}$ and we have that
\begin{eqnarray}\label{omegainst}
\omega^2= \frac{4\, c^2_{{\rm s0}} }{l_{\rm t}\,(L-l_{\rm t}) \rho_{\rm t}/\rho_0} -g\, \frac{\pi}{L},
\end{eqnarray}
where the sound speed of the background was used. According to Eq.~(\ref{omegainst}) we realise that we can have the following situations: $\omega^2>0$ ($\omega$ real), meaning that the motion is purely oscillatory around a stable equilibrium; $\omega^2<0$ ($\omega$ purely imaginary), indicating that the system is unstable; and $\omega^2=0$. This is the expected situation based on the results of Sect.~\ref{equilsect} about the equilibrium. Interestingly, it is not difficult to see that the growth rates and frequencies derived from the dispersion relation in Eq.~(\ref{omegainst}) are exactly the same $\tau$ and $\omega$ found in Sect.~\ref{sect:equil_sym} and derived from a mathematical point of view.

\begin{figure}[ht!]
        \begin{center}
        \includegraphics[trim = 0mm 0mm 0mm 0mm,clip, width=8.cm]{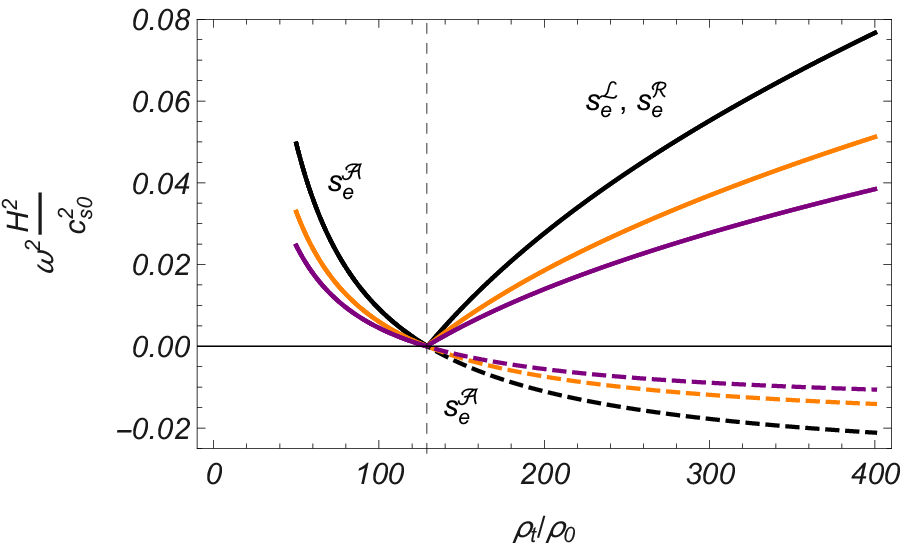}\\
        \includegraphics[trim = 0mm 0mm 0mm 0mm,clip, width=5.cm]{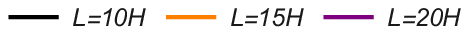}\\
        \includegraphics[trim = 0mm 0mm 0mm 0mm,clip, width=8.cm]{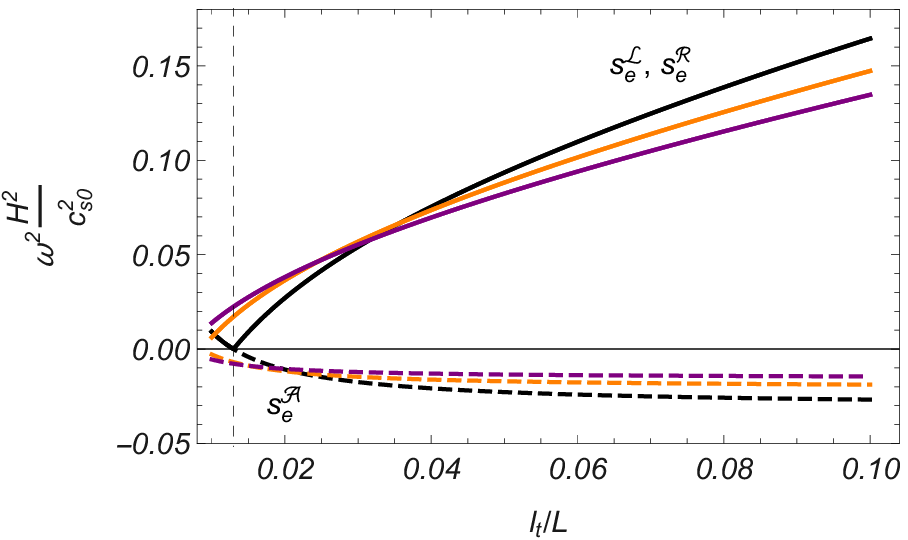}\\
        \includegraphics[trim = 0mm 0mm 0mm 0mm,clip, width=5.cm]{legendL.eps}\\
        \includegraphics[trim = 0mm 0mm 0mm 0mm,clip, width=8.cm]{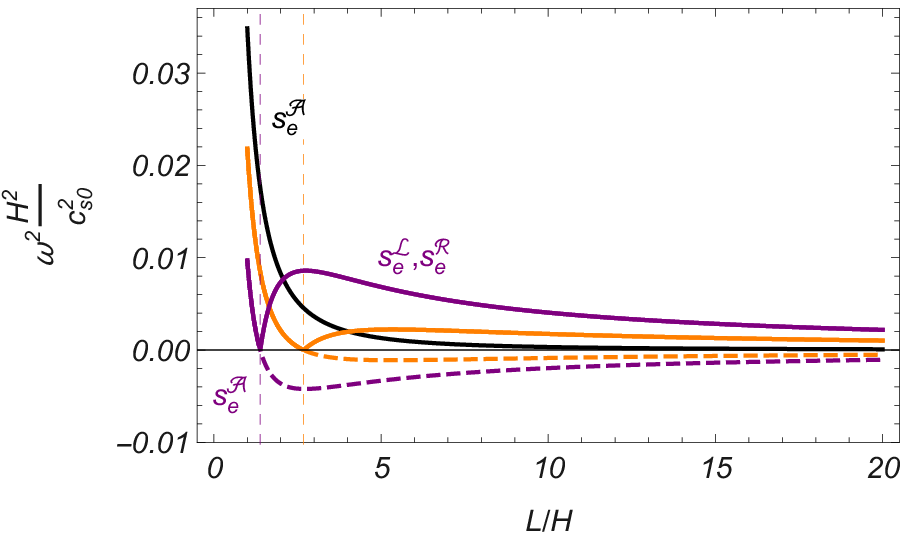}\\
        \includegraphics[trim = 0mm 0mm 0mm 0mm,clip, width=6.cm]{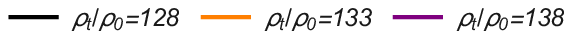}
        \caption{$\omega^2$ as a function of $\rho_{\rm t}/\rho_0$ (top panel), $l_{\rm t}/L$  (middle panel),  and $L/H$ (bottom panel) for $s_{\rm e}=s_{\rm e}^\mathscr{A}$ (below bifurcation points and lower branch curves, Eq.~(\ref{omegainst})) and $s_{\rm e}=s_{\rm e}^\mathscr{L}=s_{\rm e}^\mathscr{R}$ (upper branch curves; Eq.~(\ref{omegainstcurv})). In these plots $s_0=L/2$.  $l_{\rm t}=0.1\,H$ in the top and bottom panels and $\rho_{\rm t}/\rho_0=100$ in the middle panel. The bifurcation values (thin dashed vertical lines) have been calculated using the analytical expressions given by Eqs.~(\ref{rhoc}), (\ref{l_t}), and (\ref{lc}).} 
        \label{fig:frequency}
\end{center}
\end{figure}

The transition between the two regimes ($\omega^2=0$) provides helpful information because it corresponds to a bifurcation point, where, as we have seen, the number of solutions and/or their stability can change. This transition takes place when the following condition for the density contrast is satisfied: \begin{eqnarray}\label{rhoc}
{\rho_{\rm t}}_{\rm b}/\rho_0=\frac{4\, c^2_{{\rm s0}} \, L }{g \pi \, l_{\rm t}\, (L-l_{\rm t})},
\end{eqnarray}
where the density contrasts below this bifurcation value correspond to a stable solution, whereas for $\rho_{\rm t}>{\rho_{\rm t}}_{\rm b}$ the stable solution turns unstable. This behaviour is shown in Fig.~\ref{fig:frequency} (top panel), in which  $\omega^2$ is plotted as a function of $\rho_{\rm t}/\rho_0$ for different values of $L$ and for the three possible equilibrium positions $s_{\rm e}^\mathscr{A}$, $s_{\rm e}^\mathscr{L}$, and $s_{\rm e}^\mathscr{R}$. Since the initial position of the thread is at the apex of the structure, $s_{\rm e}^\mathscr{L}$ and $s_{\rm e}^\mathscr{R}$ are symmetric and  have the same frequency, so only one solid curve with $\omega^2>0$ is visible to the right of the bifurcation point. Frequencies associated with $s_{\rm e}^\mathscr{A}$ are found at the right of the bifurcation point and below zero in Fig.~\ref{fig:frequency}. We see that the bifurcation density, which delimits stable solutions (solid curves) with unstable (dashed curves), does not strongly depend on $L$. This behaviour agrees with Eq.~(\ref{rhoc}), which simplifies to ${\rho_{\rm t}}_{\rm b}/\rho_0=\frac{4\, c^2_{{\rm s0}} \,}{g \pi \, l_{\rm t}}$ when $l_{\rm t}$ is neglected with respect to $L$.

Assuming that $\rho_{\rm t}/\rho_0$ and $L$ are fixed, then we have a bifurcation length for the thread that determines if the solution is stable or unstable as follows:
\begin{eqnarray}\label{l_t}
{l_{\rm t}}_{\rm b}=\frac{L}{2}\left(1-\sqrt{1-\frac{16\, c^2_{{\rm s0}} }{g \pi \, L \rho_{\rm t}/\rho_0}}\right),
\end{eqnarray}
(see middle panel of Fig.~\ref{fig:frequency}). For typical prominence values the term inside the square root is always positive and the square root of this term is smaller than 1, providing a physically acceptable value (${l_{\rm t}}_{\rm b}>0$). The equilibrium positions of threads with lengths shorter than ${l_{\rm t}}_{\rm b}$ are always stable. 

Finally, from  Eq.~(\ref{omegainst}) we find that the bifurcation length of the magnetic field lines is given by
\begin{eqnarray}\label{lc}
L_{\rm b} = \frac{l_{\rm t}}{{1-\frac{4\, c^2_{{\rm s0}} }{g \pi \, l_{\rm t}\, \rho_{\rm t}/\rho_0}}},
\end{eqnarray}
and the term in the denominator is not necessarily positive. A negative bifurcation length is non-physical, meaning that in this situation it is not possible to find a transition from the stable to the unstable regime. Therefore, when 
\begin{eqnarray}\label{lcrit}
\frac{4\, c^2_{{\rm s0}} }{g \pi \, l_{\rm t}\, \rho_{\rm t}/\rho_0}>1,
\end{eqnarray}
the bifurcation length is negative, thereby implying in our case that the solution is always stable ($\omega^2>0$) independent of the length $L$; this is easy to check introducing the condition given by Eq.~(\ref{lcrit}) into Eq.~(\ref{omegainst}). Figure~\ref{fig:frequency} (bottom panel) shows $\omega^2$ as a function of $L$ for different values of $\rho_{\rm t}$ and we find a case in which  Eq.~(\ref{lcrit}) is satisfied and  $\omega^2$ is always positive.
For values of the parameters that make the left-hand side of Eq.~(\ref{lcrit}) less than 1, the solution is either stable or unstable, depending on the length of the field lines, if  $L<L_{\rm b}$ the system that only develops $s_{\rm e}^\mathscr{A}$ is stable and if $L>L_{\rm b}$ it is unstable (see lower branch curves in bottom panel of Fig.~\ref{fig:frequency}). Interestingly, the bifurcation length $L_{\rm b}$ depends strongly on the density contrast (see the short range of $\rho_{\rm t}/\rho_0$ in Fig.~\ref{fig:frequency} bottom panel). We note that to have a stable solution the three criteria must be simultaneously satisfied; these criteria are $\rho_{\rm t}<{\rho_{\rm t}}_{\rm b}$, $l_{\rm t}<{l_{\rm t}}_{\rm b}$, and $L<L_{\rm b}$  or equivalently Eq.~(\ref{lcrit}) for the length of the field lines. This is an interesting result that provides useful information about the behaviour of the system depending on the parameters.

It is worthy noting that the first term on the right-hand side of Eq.~(\ref{omegainst}) is the frequency squared of a slow mode in a slab model of length $l_{\rm t}$ inside a field line with a length of $L$ \citep{joarderroberts92,diazetal10,soleretal10}. \citet{lunaetal12} derived a similar expression for the frequency but for a magnetic field with a dip at the location of the thread. In their formula the sign in front of the gravitational term is positive and therefore the instability is not present in their configuration; their system is always (linearly) stable because they consider a concave upwards geometry for the magnetic field.

Now we generalise the frequency of oscillation for any  equilibrium position, $s_{\rm e}$, along the tube and hence we do not restrict the analysis to the equilibrium solution at the tube apex. But we note that the thread is still initially located at $s=L/2$. Under these conditions we have that
\begin{eqnarray}\label{adiabaticwavem}
p_1(\delta s)&=&p_{10} \left(\frac{s_{\rm e}-l_{\rm t}/2}{s_{\rm e}-l_{\rm t}/2+\delta s}\right)^\gamma,\nonumber\\ 
p_2( \delta s)&=&p_{20} \left(\frac{s_{\rm e}-l_{\rm t}/2}{L-s_{\rm e}-l_{\rm t}/2-\delta s}\right)^\gamma,
\end{eqnarray}
where $p_{10}$ and $p_{20}$ correspond to the equilibrium pressures that satisfy Eq.~(\ref{adiabatic1}). We rewrite ${p_1}$ as 
\begin{eqnarray}\label{adiabaticwave1m}
{p_1(\delta s)}\simeq  p_{10} \left[1- \gamma \delta s/(s_{\rm e}-l_{\rm t}/2)\right],
\end{eqnarray}
where  we used the Maclaurin series approximation for small arguments, but  if $s_{\rm e}$ is similar to $l_{\rm t}/2$ the approximation is not valid. An analogous expression is derived for ${p_2}$. 
The pressure gradient using the previous equations is approximated by
\begin{eqnarray}\label{aproxpresswavec}
\frac{dp}{ds}(\delta s) &\simeq& \frac{p_2(\delta s)-p_1( \delta s)}{l_{\rm t}}\nonumber\\&=& \frac{1}{l_{\rm t}}\left(p_{20}-p_{10} + \frac{p_{20}\, \gamma\, \delta s}{L-s_{\rm e}-l_{\rm t}/2}+ \frac{p_{10}\, \gamma\, \delta s}{s_{\rm e}-l_{\rm t}/2}\right).\nonumber\\
\end{eqnarray}
For the projected gravity we need to perform a Taylor expansion of the cosine around the equilibrium position, which is now $s_{\rm e}$,
\begin{eqnarray}\label{cosexpand}
\cos\left(\pi \frac{s}{L}\right)=\cos\left(\pi \frac{s_{\rm e}}{L}\right)-\frac{\pi}{L}\sin\left(\pi \frac{s_{\rm e}}{L}\right)(s-s_{\rm e})+...,
\end{eqnarray}
where we only retain up to linear terms in $s-s_{\rm e}$. We have that $\delta s=s-s_{\rm e}$ according to our notation. 

In the momentum equation, using Eqs.~(\ref{aproxpresswavec}) and (\ref{cosexpand}), we find two terms that are not proportional to $\delta s$ but that cancel out because they correspond to the equilibrium condition given by Eq.~(\ref{eqequil}). We derive that the square of the frequency is
given by\begin{eqnarray}\label{omegainstcurv}
\omega^2&=&\frac{c^2_{{\rm s0}}}{l_{\rm t} \rho_{\rm t}/\rho_0}\left[\frac{\left(L/2-l_{\rm t}/2\right)^\gamma}{(L-s_{\rm e}-l_{\rm t}/2)^{\gamma+1}}+\frac{\left(L/2-l_{\rm t}/2\right)^\gamma}{(s_{\rm e}-l_{\rm t}/2)^{\gamma+1}}\right]\nonumber \\ &-&g\, \frac{\pi}{L}\sin\left(\pi \frac{s_{\rm e}}{L}\right).
\end{eqnarray}
\noindent
When $s_{\rm e}=L/2$, that is the trivial solution, Eq.~(\ref{omegainstcurv}) is exactly the same as Eq.~(\ref{omegainst}), recovering the previous situation. The solutions of Eq.~(\ref{omegainstcurv}) are shown in Fig.~\ref{fig:frequency}.  Below the bifurcation values ${\rho_{\rm t}}_{\rm b}$, ${l_{\rm t}}_{\rm b}$, and $L_{\rm b}$, the equilibrium solution is $s_{\rm e}^\mathscr{A}$. When the bifurcation values are reached, $s_{\rm e}^\mathscr{A}$ becomes unstable ($\omega^2<0$), and as expected, $s_{\rm e}^\mathscr{L}$ is stable ($\omega^2>0$). For $s_{\rm e}^\mathscr{R}$ the value of frequency is exactly the same as that for $s_{\rm e}^\mathscr{L}$ and therefore only one curve is visible in the plot. For $s_{\rm e}^\mathscr{L}$ and $s_{\rm e}^\mathscr{R}$ (upper branch curves) the frequency squared monotonically increases with $\rho_{\rm t}$ and $l_{\rm t}$, but it increases with $L$ until a maximum and then it starts decreasing with  $L$.  It is important to mention that all the results shown in Fig.~\ref{fig:frequency} are the solutions of the configurations shown in Fig.~\ref{fig:equilbifurc} and the stability analysis done in Sect.~\ref{sect:equil_sym} agrees with the results in the present section.  

\begin{figure}[ht!]
        \begin{center}
        \includegraphics[trim = 0mm 0mm 0mm 0mm,clip, width=8.cm]{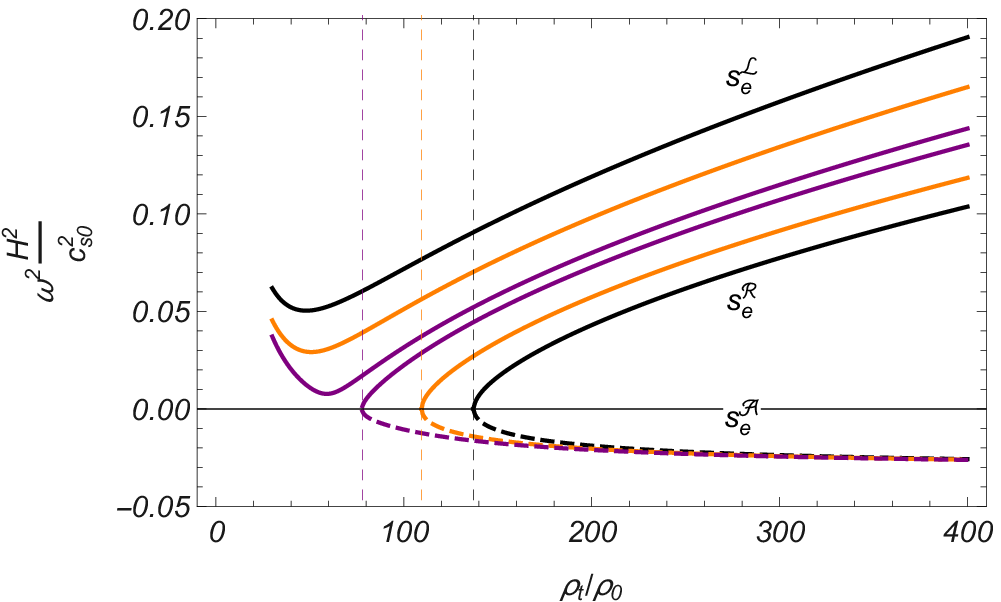}
        \includegraphics[trim = 0mm 0mm 0mm 0mm,clip, width=8.cm]{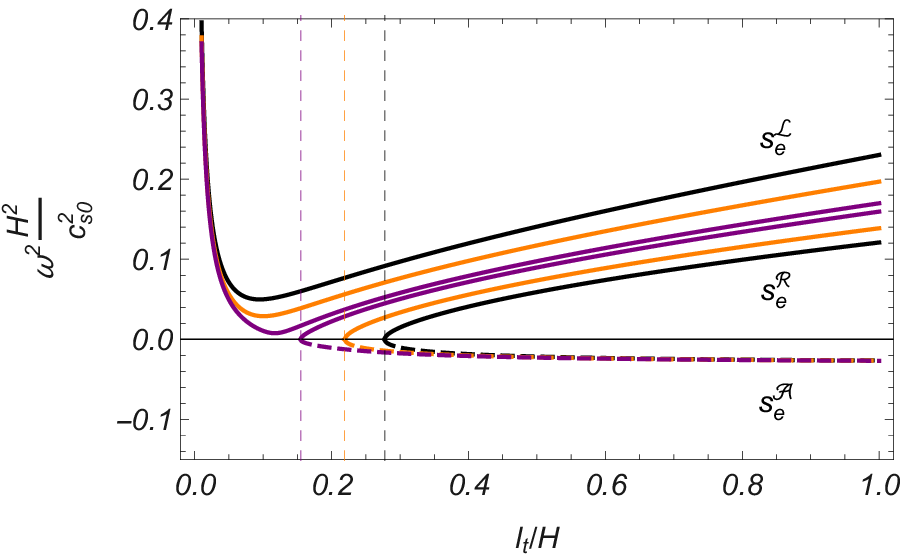}
        \includegraphics[trim = 0mm 0mm 0mm 0mm,clip, width=8.cm]{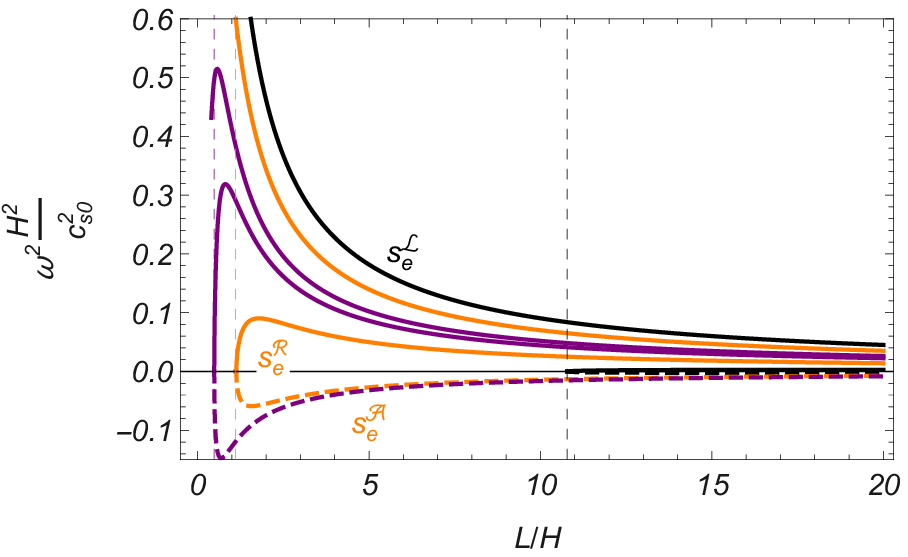}
        \includegraphics[trim = 0mm 0mm 0mm 0mm,clip, width=6.cm]{legends0.eps}
        \caption{$\omega^2$ as function of different equilibrium parameters (Eq.~(\ref{omegainstcurvfinal})) for three different initial positions of the thread, $s_0$. The expression $L=10\,H$ and $l_{\rm t}=0.2\,H$ (top panel), $\rho_{\rm t}/\rho_0=100$ and $L=10\,H$ (middle panel), and $l_{\rm t}=0.2\,H$ and $\rho_{\rm t}/\rho_0=137$ (bottom panel). The vertical lines correspond to the bifurcation points and are calculated by imposing that {$\omega^2=0$}.}
        \label{fig:w2s0}
\end{center}
\end{figure}

\subsection{Thread initially at any location along the tube}

Finally, using the same procedure as before we derive an expression for the oscillation frequency of a thread around the final equilibrium position $s_{\rm e}$, which has been achieved from an initial location of the thread at $s_0\neq L/2$. The corresponding pressures at the edge of the thread due to a small displacement $\delta s$ from the equilibrium position are expressed as Eq.~(\ref{adiabaticwavem}),
but now $p_{10}$ and $p_{20}$ correspond to the equilibrium pressures that satisfy Eq.~(\ref{adiabatic2}). Using similar approximations to those in Eq.~(\ref{adiabaticwave1m}) and cancelling the equilibrium terms according to Eq.~(\ref{eqequil2}), we eventually find that the frequency squared is written as 
\begin{eqnarray}\label{omegainstcurvfinal}
\omega^2&=&\frac{c^2_{{\rm s0}}}{l_{\rm t} \rho_{\rm t}/\rho_0}\left[\frac{\left(L-s_0-l_{\rm t}/2\right)^\gamma}{(L-s_{\rm e}-l_{\rm t}/2)^{\gamma+1}}+\frac{\left(s_0-l_{\rm t}/2\right)^\gamma}{(s_{\rm e}-l_{\rm t}/2)^{\gamma+1}}\right]\nonumber\\&-&g\, \frac{\pi}{L}\sin\left(\pi \frac{s_{\rm e}}{L}\right).
\end{eqnarray}
The previous frequency expressions given in Eq.~(\ref{omegainst}) when $s_0=s_{\rm e}=L/2$ and Eq.~(\ref{omegainstcurv}) when $s_0=L/2$, $s_{\rm e}\neq L/2$ are recovered in Eq.~(\ref{omegainstcurvfinal}). In this equation $s_{\rm e}$ must be a solution to Eq.~(\ref{eqequil2}). Obtaining analytical bifurcation values, except for the trivial case ($s_0=s_{\rm e}=L/2$), is much more difficult now and it is only possible for the bifurcation density contrast. The rest of bifurcation values are determined numerically. As expected from bifurcation theory, imposing the condition for a bifurcation value, that is $\omega^2=0$ in Eq.~(\ref{omegainstcurvfinal}), is completely equivalent to the condition given by Eq.~(\ref{eqequil2d}) that determines the bifurcation point based only on the equilibrium configuration. This is confirmed by the comparison of the bifurcation values in  Fig.~\ref{fig:w2s0} and the respective results in Fig.~\ref{fig:equilbifurcnosim}. This is also a solid indication that the performed eigenfrequency analysis in this section is coherent. 

\begin{figure}[ht!]
        \begin{center}
        \includegraphics[trim = 0mm 0mm 0mm 0mm,clip, width=8.cm]{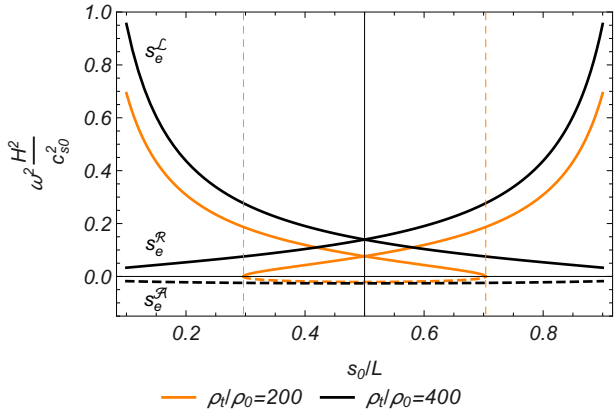}\\
                \includegraphics[trim = 0mm 0mm 0mm 0mm,clip, width=4.cm]{legendrho1.eps}
        \caption{$\omega^2$ as a function of $s_0/L$ (Eq.~(\ref{omegainstcurvfinal})) for the three solutions obtained in Fig.~\ref{fig:se_s0_two}. The solid lines correspond to $s_{\rm e}^\mathscr{L}$ (lines decreasing from left to right) and $s_{\rm e}^\mathscr{R}$ (lines increasing from left to right), and the dashed lines to $s_{\rm e}^\mathscr{A}$. The horizontal line represents $\omega=0$ and divides the plot into stable and unstable solutions. In this plot the left and right solutions are stable while the near apex  solution is unstable. The bifurcation points (vertical dashed lines) have been calculated by solving Eqs.~(\ref{eqequil2}) and (\ref{eqequil2d}) simultaneously, and it is equivalent to impose $\omega^2=0$ in Eq.~(\ref{omegainstcurvfinal}) once $s_{\rm e}$ is known.}
        \label{fig:w2_s0_two}
\end{center}
\end{figure}

Now the break of symmetry due to the imperfection parameter $s_0$ causes the difference between the frequency squared of $s_{\rm e}^\mathscr{L}$ and $s_{\rm e}^\mathscr{R}$. This is visible in the plotted  upper curves of Fig.~\ref{fig:w2s0} (solid curves). Beyond the bifurcation values two branches appear that correspond to $s_{\rm e}^\mathscr{R}$ and $s_{\rm e}^\mathscr{A}$, which are positive for $s_{\rm e}^\mathscr{R}$ (solid curves) and negative for $s_{\rm e}^\mathscr{A}$ (dashed curves). Logically, when $s_0/L\rightarrow 0.5$ (purple curves), $\omega^2$ for $s_{\rm e}^\mathscr{R}$ approaches the same value as for $s_{\rm e}^\mathscr{L}$. For $s_{\rm e}^\mathscr{R}$ (emerging at the bifurcation point), $\omega^2$ increases with $\rho_{\rm t}/\rho_0$ (top panel) and $l_{\rm t}/H$ (middle panel). For $s_{\rm e}^\mathscr{L}$, stable in the whole parameter range, $\omega^2$ decreases with $\rho_{\rm t}/\rho_0$ and $l_{\rm t}/H$ until it reaches a minimum located before the bifurcation point from where it starts increasing. On the contrary, in the bottom panel of Fig.~\ref{fig:w2s0} we see that for $s_{\rm e}^\mathscr{L}$ the frequency decreases for large values of $L$.

In Fig.~\ref{fig:w2_s0_two} the three square of the frequencies corresponding to the solutions in Fig.~\ref{fig:se_s0_two} are plotted as a function of the initial position of the thread, $s_0$. To calculate these frequencies, we used the values of $s_{\rm e}$ that satisfy Eq.~(\ref{eqequil2}). We note that $s_{\rm e}^{\mathscr{L}}$ and $s_{\rm e}^{\mathscr{R}}$ are always stable while $s_{\rm e}^\mathscr{A}$ is always unstable. As expected from physical grounds the left and right side solutions converge to the same value when there is symmetry in the system, that is for $s_0=L/2$; we note that beyond this point the meaning of $s_{\rm e}^{\mathscr{L}}$ and $s_{\rm e}^{\mathscr{R}}$ has to be inverted. We also observe the transition between one and three solutions for a specific choice of parameters ($\rho_{\rm t}/\rho_0=200$, orange curves). At the bifurcation point two solutions emerge from $\omega=0$: one representing a stable solution ($s_{\rm e}^\mathscr{R}$) and another corresponding to an unstable solution ($s_{\rm e}^\mathscr{A}$). We also see that $\omega^2$ significantly varies with $s_0/L$ for stable solutions, but it is essentially independent of $s_0/L$ for unstable solutions. 

\section{Conclusions and discussion}\label{conclusions}

The thread-like structures studied in this work appear to be present in filaments everywhere \citep{linetal2008}, along the horizontal spines or the quasi-vertical barbs. High-resolution observations of the solar corona show that the fine structure of prominences is very dynamic. Even during the prominence formation we observe plasma flows along the magnetic fields. The localised heating above the flux tube footpoints produces a cyclic pattern of evaporation of the chromospheric plasma, which condenses in the coronal part of the tube and produces the plasma flows. During the process, the pressure imbalance between the chromosphere and the condensation region would push the chromospheric plasma upwards to supply mass to the corona in the form of siphon flows \citep{xiaetal2011,zhouetal2014}. In spite of the complicated dynamics of prominence threads, it is interesting to perform the stability analysis of these structures. For this reason, we developed a very idealised model that provides the basic physics to understand the evolution of a density enhancement or thread that moves along a curved concave magnetic field, a very common situation in the solar corona. The main assumption in the model is that the behaviour of the thread is essentially that of a piston that in our case is governed by gas pressure and gravity. Our model facilitates the existence of a suspended mass in the absence of magnetic dips.

We derived two equations for the equilibrium  of a thread initially located at any position along the tube (Eq.~(\ref{eqequil2})) and the corresponding eigenfrequency of oscillation (Eq.~(\ref{omegainstcurvfinal})), which revealed useful knowledge about the stability properties of the system. In spite of the simplicity of our model the equilibrium equation is very rich from a mathematical point of view and leads to significant physical interpretations. Different scenarios are possible; a density enhancement initially located at the apex is always in equilibrium in our model, but this equilibrium is either stable or unstable. We derived analytical expressions for the bifurcation values of the different parameters when a transition between a stable and unstable regime is produced. If the thread near to the apex is unstable then the system allows two additional symmetric equilibria at lower heights that are stable. Under these circumstances a thread situated at the apex of the tube moves along the magnetic field until it reaches the lower state energy associated with the new equilibrium. When the density enhancement is not initially at the apex we have in essence a similar behaviour: a single stable solution or three solutions, two of which are stable and the other unstable. In this last case, the mathematical behaviour of the system is characterised by the appearance of a S-shaped bifurcation, well known in bifurcation theory.
 
Our analysis reveals the presence of bifurcation points, obtained analytically when the thread is initially located at the apex. For example, when the condition given by Eq.~(\ref{lcrit}) is satisfied, the system is stable, independent of the length of the magnetic field lines where the thread is suspended. Otherwise, a thread with a fixed density contrast and length can be stable when it is located along short magnetic field lines, but unstable on long magnetic field lines, namely, higher up in the magnetic arcade. This can have some implications regarding the finite height of observed solar prominences, but also concerning the stability of unbounded models with height such as those of \citet{hoodanzer90}. However, the presence of magnetic dips, missing in our model, can have important consequences regarding stability.

The length of the magnetic tubes plays a key role in the equilibrium and stability of the threads. The physical interpretation of the dependence of the instability on the length of the field lines can be  connected with the pressure-volume work. The apex location for threads in short flux tubes is more stable because, with the same perturbation, the gas pressure increases more drastically in the compressed region with the decreasing volume when the initial gas volume is small (and decreases more rapidly in the expanding region).

A natural extension of the model presented in this work is the inclusion of magnetic dips through the straightforward modification of the projected gravity along the field. This most likely has relevant consequences regarding stability and affects the eventual bifurcation points that may appear in the system. This work is in progress and is addressed in part \rm {II} as a continuation of the present paper part \rm {I}. Another future improvement to the model is the implementation of the variation of the cross-sectional area along the tube, allowing us to incorporate a truly 3D effect in the calculation of the equilibrium and the corresponding eigenfrequencies \citep[see][]{lunaetal16_2}.

\citet{an1988} and \citet{wu1990} investigated the effects of plasma injection on the formation of the Kippenhahn-Schl\"uter model of prominence in optimum conditions. These authors found that for high values of the plasma-$\beta$ parameter (the ratio of plasma pressure to magnetic pressure) the magnetic arcade develops a magnetic dip at the centre of the structure that supports the prominence plasma. However, comparing with our study, in the low plasma-$\beta$ regime (or under others injection conditions) they found that the dip is less deep and the system develops two additional plasma enhancements located at the lateral edges of the magnetic arcade. Recent works suggest that the deformation of the magnetic field lines is determined by the parameter $\delta$ (the ratio of the gravity to the magnetic pressure) \citep{zhouetal2018,zhangetal2019}. \citet{an1988} suggested that the steady lateral plasma accumulates because of both the injection process and because the field lines without dips do not geometrically contain the injected plasma, but \citet{wu1990} proposed that the prominence mass is also supported by an increase in the pressure gradient. Since in this study we consider that magnetic field lines do not change owing to the presence of the dense thread we investigate in detail the results of \citet{an1988} and \citet{wu1990} in  part \rm {II} of this paper. On the contrary, studies of the formation of 1D filament threads by chromospheric heating in the presence of non-adiabatic effects, such as radiative losses and thermal conduction, show that for magnetic loops without a dip, the plasma condenses but it streams along the magnetic field and disappears after falling to the footpoints \citep{antiochosetal2000,karpenetal2006}. Moreover, when the thread is initially in a thermal and force-balance equilibrium state  but it is disturbed by a strong velocity perturbation, the prominence mass drains down to the chromosphere. Dense blobs of falling plasma have been habitually observed \citep{SchrijverandCarolus2001,groofetal2005}, therefore it seems that threads cannot be held static along vertical magnetic flux tubes in the corona. The coronal part of the tube can only slow down the falling blobs. \citet{mulleretal2004} proposed that the acceleration reduces because the pressure of the cooling plasma underneath the radiating blobs slows down the descent, and \citet{2014ApJ...784...21O,2016ApJ...818..128O} and \citet{martinez-gomezetal2020} argue that pressure gradient is the main force that opposes the action of gravity. Our study proposes that the pressure gradient can cause the equilibrium of threads in quasi-vertical flux tubes without dips even though it has not been corroborated by observations. Besides, our model is relatively simple to study the stability of filament threads in the magnetic field without dips. Other processes such as radiation and heat conduction must be considered, which might change the stability results.    

Our model also demonstrates that when the mass of the thread is high enough the position of the equilibrium is near to the footpoints. This allows us to constrain the parameters to find solutions that do not represent suspended threads but correspond to threads that travel down the tube until they essentially settle down into a new equilibrium around the base of the corona. This last situation is applicable, for example, to coronal rain falling along magnetic field lines. In this regard, our study can be relevant to explain some results of 2D and 2.5D numerical simulations of dense blobs released in the solar atmosphere \citep[e.g.][]{macga2001,kohutovaver2017a,kohutovaver2017b}. These simulations have, in common with our work, a dense blob (akin to our thread) that is released in a gravitationally stratified background atmosphere with a closed bottom boundary at the base of the corona. This set-up allows the dense material falling along the magnetic tube to act as a piston. For this reason, \citet{macga2001} and \citet{kohutovaver2017a,kohutovaver2017b} find, as we do, that when the blob density is large enough this cold material falls down to the surface because the increased pressure that builds underneath it cannot counteract the blob weight. On the contrary, blobs with smaller mass can find an equilibrium position before reaching the surface and so they oscillate around this position. In addition to the relevance of the blob density, $\rho_{\rm t}$, we also note that the length of magnetic field lines and the blob length are important parameters that control the system dynamics.

\begin{acknowledgements} A. A., J. T., R. O., and M. C. acknowledge the support from grant AYA2017-85465-P
(MINECO/AEI/FEDER, UE), to the Conselleria d'Innovaci\'o, Recerca i Turisme del
Govern Balear, and also to IAC$^3$. A.A. acknowledges the Spanish `Ministerio de Econom\'ia, Industria y Competitividad’ for the `Ayuda para contratos predoctorales’ grant BES-2015-075040.  We are grateful to Rafel Prohens from the Departament de Ci\`encies Matem\`atiques i Inform\`atica, Universitat de les Illes Balears (UIB) for his advise on bifurcation theory. All the numerical calculations in this work have been performed using Mathematica \citep{Mathematica}.
\end{acknowledgements}

\bibliographystyle{aa}

%\bibliography{ref} 

\begin{thebibliography}{43}
	\expandafter\ifx\csname natexlab\endcsname\relax\def\natexlab#1{#1}\fi
	
	\bibitem[{{Adrover-Gonz{\'a}lez} \& {Terradas}(2020)}]{adroverterradas2020}
	{Adrover-Gonz{\'a}lez}, A. \& {Terradas}, J. 2020, \aap, 633, A113
	
	\bibitem[{{An} {et~al.}(1988){An}, {Bao}, {Wu}, \& {Suess}}]{an1988}
	{An}, C.~H., {Bao}, J.~J., {Wu}, S.~T., \& {Suess}, S.~T. 1988, \solphys, 115,
	93
	
	\bibitem[{{Antiochos} {et~al.}(2000){Antiochos}, {MacNeice}, \&
		{Spicer}}]{antiochosetal2000}
	{Antiochos}, S.~K., {MacNeice}, P.~J., \& {Spicer}, D.~S. 2000, \apj, 536, 494
	
	\bibitem[{{Blokland} \& {Keppens}(2011)}]{bloklandkeppens2011}
	{Blokland}, J.~W.~S. \& {Keppens}, R. 2011, \aap, 532, A93
	
	\bibitem[{{de Bruyne} \& {Hood}(1993)}]{debruynehood93}
	{de Bruyne}, P. \& {Hood}, A.~W. 1993, \solphys, 147, 97
	
	\bibitem[{{de Groof} {et~al.}(2005){de Groof}, {Bastiaensen}, {M{\"u}ller},
		{Berghmans}, \& {Poedts}}]{groofetal2005}
	{de Groof}, A., {Bastiaensen}, C., {M{\"u}ller}, D.~A.~N., {Berghmans}, D., \&
	{Poedts}, S. 2005, \aap, 443, 319
	
	\bibitem[{{D{\'{\i}}az} {et~al.}(2010){D{\'{\i}}az}, {Oliver}, \&
		{Ballester}}]{diazetal10}
	{D{\'{\i}}az}, A.~J., {Oliver}, R., \& {Ballester}, J.~L. 2010, \apj, 725, 1742
	
	\bibitem[{{Engvold}(2015)}]{engvold15}
	{Engvold}, O. 2015, Astrophysics and Space Science Library, Vol. 415,
	{Description and Classification of Prominences} (Springer, Cham), 31
	
	\bibitem[{{Fiedler} \& {Hood}(1992)}]{fiedlerhood1992}
	{Fiedler}, R.~A.~S. \& {Hood}, A.~W. 1992, \solphys, 141, 75
	
	\bibitem[{{Hillier} \& {van Ballegooijen}(2013)}]{hilliervan2013}
	{Hillier}, A. \& {van Ballegooijen}, A. 2013, \apj, 766, 126
	
	\bibitem[{{Hood} \& {Anzer}(1990)}]{hoodanzer90}
	{Hood}, A.~W. \& {Anzer}, U. 1990, \solphys, 126, 117
	
	\bibitem[{{Joarder} \& {Roberts}(1992)}]{joarderroberts92}
	{Joarder}, P.~S. \& {Roberts}, B. 1992, \aap, 261, 625
	
	\bibitem[{Jordan \& Smith(1987)}]{jordansmith1987}
	Jordan, D.~W. \& Smith, P. 1987, Nonlinear Ordinary Differential Equations (2nd
	Ed.) (USA: Oxford University Press, Inc.)
	
	\bibitem[{{Karpen} {et~al.}(2006){Karpen}, {Antiochos}, \&
		{Klimchuk}}]{karpenetal2006}
	{Karpen}, J.~T., {Antiochos}, S.~K., \& {Klimchuk}, J.~A. 2006, \apj, 637, 531
	
	\bibitem[{{Kohutova} \& {Verwichte}(2017{\natexlab{a}})}]{kohutovaver2017a}
	{Kohutova}, P. \& {Verwichte}, E. 2017{\natexlab{a}}, \aap, 602, A23
	
	\bibitem[{{Kohutova} \& {Verwichte}(2017{\natexlab{b}})}]{kohutovaver2017b}
	{Kohutova}, P. \& {Verwichte}, E. 2017{\natexlab{b}}, \aap, 606, A120
	
	\bibitem[{{Lin} {et~al.}(2008){Lin}, {Martin}, \& {Engvold}}]{linetal2008}
	{Lin}, Y., {Martin}, S.~F., \& {Engvold}, O. 2008, in Astronomical Society of
	the Pacific Conference Series, Vol. 383, Subsurface and Atmospheric
	Influences on Solar Activity, ed. R.~{Howe}, R.~W. {Komm}, K.~S.
	{Balasubramaniam}, \& G.~J.~D. {Petrie}, 235
	
	\bibitem[{{Low} \& {Zhang}(2004)}]{lowzhang2004}
	{Low}, B.~C. \& {Zhang}, M. 2004, \apj, 609, 1098
	
	\bibitem[{{Luna} {et~al.}(2012){Luna}, {D{\'{\i}}az}, \& {Karpen}}]{lunaetal12}
	{Luna}, M., {D{\'{\i}}az}, A.~J., \& {Karpen}, J. 2012, \apj, 757, 98
	
	\bibitem[{{Luna} {et~al.}(2016{\natexlab{a}}){Luna}, {D{\'\i}az}, {Oliver},
		{Terradas}, \& {Karpen}}]{lunaetal16_2}
	{Luna}, M., {D{\'\i}az}, A.~J., {Oliver}, R., {Terradas}, J., \& {Karpen}, J.
	2016{\natexlab{a}}, \aap, 593, A64
	
	\bibitem[{{Luna} {et~al.}(2016{\natexlab{b}}){Luna}, {Terradas}, {Khomenko},
		{Collados}, \& {de Vicente}}]{lunaetal2016}
	{Luna}, M., {Terradas}, J., {Khomenko}, E., {Collados}, M., \& {de Vicente}, A.
	2016{\natexlab{b}}, \apj, 817, 157
	
	\bibitem[{{Mackay} \& {Galsgaard}(2001)}]{macga2001}
	{Mackay}, D.~H. \& {Galsgaard}, K. 2001, \solphys, 198, 289
	
	\bibitem[{{Mackay} {et~al.}(2010){Mackay}, {Karpen}, {Ballester}, {Schmieder},
		\& {Aulanier}}]{mackayetal10}
	{Mackay}, D.~H., {Karpen}, J.~T., {Ballester}, J.~L., {Schmieder}, B., \&
	{Aulanier}, G. 2010, \ssr, 151, 333
	
	\bibitem[{{Mart{\'\i}nez-G{\'o}mez} {et~al.}(2020){Mart{\'\i}nez-G{\'o}mez},
		{Oliver}, {Khomenko}, \& {Collados}}]{martinez-gomezetal2020}
	{Mart{\'\i}nez-G{\'o}mez}, D., {Oliver}, R., {Khomenko}, E., \& {Collados}, M.
	2020, \aap, 634, A36
	
	\bibitem[{{M{\"u}ller} {et~al.}(2004){M{\"u}ller}, {Peter}, \&
		{Hansteen}}]{mulleretal2004}
	{M{\"u}ller}, D.~A.~N., {Peter}, H., \& {Hansteen}, V.~H. 2004, \aap, 424, 289
	
	\bibitem[{{Oliver} {et~al.}(2016){Oliver}, {Soler}, {Terradas}, \&
		{Zaqarashvili}}]{2016ApJ...818..128O}
	{Oliver}, R., {Soler}, R., {Terradas}, J., \& {Zaqarashvili}, T.~V. 2016, \apj,
	818, 128
	
	\bibitem[{{Oliver} {et~al.}(2014){Oliver}, {Soler}, {Terradas}, {Zaqarashvili},
		\& {Khodachenko}}]{2014ApJ...784...21O}
	{Oliver}, R., {Soler}, R., {Terradas}, J., {Zaqarashvili}, T.~V., \&
	{Khodachenko}, M.~L. 2014, \apj, 784, 21
	
	\bibitem[{{Parker}(1979)}]{parker1979}
	{Parker}, E.~N. 1979, {Cosmical magnetic fields. Their origin and their
		activity} (Clarendon Press)
	
	\bibitem[{{Petrie} {et~al.}(2007){Petrie}, {Blokland}, \&
		{Keppens}}]{petrietal2007}
	{Petrie}, G.~J.~D., {Blokland}, J.~W.~S., \& {Keppens}, R. 2007, \apj, 665, 830
	
	\bibitem[{{Schrijver}(2001)}]{SchrijverandCarolus2001}
	{Schrijver}, C.~J. 2001, \solphys, 198, 325
	
	\bibitem[{{Soler} {et~al.}(2010){Soler}, {Arregui}, {Oliver}, \&
		{Ballester}}]{soleretal10}
	{Soler}, R., {Arregui}, I., {Oliver}, R., \& {Ballester}, J.~L. 2010, \apj,
	722, 1778
	
	\bibitem[{Strogatz(2018)}]{strogatz2018}
	Strogatz, S. 2018, Nonlinear Dynamics and Chaos: With Applications to Physics,
	Biology, Chemistry, and Engineering (CRC Press)
	
	\bibitem[{{Terradas} {et~al.}(2013){Terradas}, {Soler}, {D{\'\i}az}, {Oliver},
		\& {Ballester}}]{terradasetal2013}
	{Terradas}, J., {Soler}, R., {D{\'\i}az}, A.~J., {Oliver}, R., \& {Ballester},
	J.~L. 2013, \apj, 778, 49
	
	\bibitem[{{Terradas} {et~al.}(2015){Terradas}, {Soler}, {Luna}, {Oliver}, \&
		{Ballester}}]{terradasetal2015}
	{Terradas}, J., {Soler}, R., {Luna}, M., {Oliver}, R., \& {Ballester}, J.~L.
	2015, \apj, 799, 94
	
	\bibitem[{{Terradas} {et~al.}(2016){Terradas}, {Soler}, {Luna}, {Oliver},
		{Ballester}, \& {Wright}}]{terradasetal2016}
	{Terradas}, J., {Soler}, R., {Luna}, M., {et~al.} 2016, \apj, 820, 125
	
	\bibitem[{Wang(1994)}]{wang1994}
	Wang, S.-H. 1994, Nonlinear Analysis: Theory, Methods \& Applications, 22, 1475
	
	\bibitem[{{Wiggins}(2003)}]{wiggins2003}
	{Wiggins}, S. 2003, {Introduction to Applied Nonlinear Dynamical Systems and
		Chaos } (Springer-Verlag New York)
	
	\bibitem[{Wolfram~Research(2020)}]{Mathematica}
	Wolfram~Research, I. 2020, Mathematica, {V}ersion 12.1, champaign, IL, 2020
	
	\bibitem[{{Wu} {et~al.}(1990){Wu}, {Bao}, {An}, \& {Tandberg-Hanssen}}]{wu1990}
	{Wu}, S.~T., {Bao}, J.~J., {An}, C.~H., \& {Tandberg-Hanssen}, E. 1990,
	\solphys, 125, 277
	
	\bibitem[{{Xia} {et~al.}(2011){Xia}, {Chen}, {Keppens}, \& {van
			Marle}}]{xiaetal2011}
	{Xia}, C., {Chen}, P.~F., {Keppens}, R., \& {van Marle}, A.~J. 2011, \apj, 737,
	27
	
	\bibitem[{{Zhang} {et~al.}(2019){Zhang}, {Fang}, \& {Chen}}]{zhangetal2019}
	{Zhang}, L.~Y., {Fang}, C., \& {Chen}, P.~F. 2019, \apj, 884, 74
	
	\bibitem[{{Zhou} {et~al.}(2014){Zhou}, {Chen}, {Zhang}, \&
		{Fang}}]{zhouetal2014}
	{Zhou}, Y.-H., {Chen}, P.-F., {Zhang}, Q.-M., \& {Fang}, C. 2014, Research in
	Astronomy and Astrophysics, 14, 581
	
	\bibitem[{{Zhou} {et~al.}(2018){Zhou}, {Xia}, {Keppens}, {Fang}, \&
		{Chen}}]{zhouetal2018}
	{Zhou}, Y.-H., {Xia}, C., {Keppens}, R., {Fang}, C., \& {Chen}, P.~F. 2018,
	\apj, 856, 179
	
\end{thebibliography}

\end{document}